\documentclass[fleqn,10pt]{wlscirep}
\usepackage[utf8]{inputenc}
\usepackage[T1]{fontenc}

\usepackage{longtable}

\def\apj{Astrophys.~J.}
\def\apjl{Astrophys.~J.~Lett.}
\def\apjs{Astrophys.~J.~Suppl.}
\def\mnras{Mon.~Not.~R.~Astron.~Soc.}

\def\nat{Nature}

\def\pasp{Publ.~Astron.~Soc.~Pac.}

\def\aap{Astron.~Astrophys.}
\def\araa{Annu.~Rev.~Astron.~Astrophys.}

\def\apss{Astron. Space Sci.}
\def\ssr{Space Sci. Rev.}

\title{A Superluminous Supernova Lightened by Collisions with Pulsational Pair-instability Shells}

\author[1]{Weili Lin}
\author[1,2,3,*]{Xiaofeng Wang}
\author[4]{Lin Yan}
\author[5]{Avishay Gal-Yam}
\author[1]{Jun Mo}
\author[6]{Thomas G. Brink}
\author[6]{Alexei V. Filippenko}
\author[1]{Danfeng Xiang}
\author[7]{Ragnhild Lunnan}
\author[6]{Weikang Zheng}
\author[8,9]{Peter Brown}
\author[10]{Mansi Kasliwal}
\author[10]{Christoffer Fremling}
\author[11,12,13]{Nadejda Blagorodnova}
\author[14]{Davron Mirzaqulov}
\author[14,15]{Shuhrat A. Ehgamberdiev}
\author[16,17,1]{Han Lin}
\author[1]{Kaicheng Zhang}
\author[18]{Jicheng Zhang}
\author[1]{Shengyu Yan}
\author[16,17,19]{Jujia Zhang}
\author[1]{Zhihao Chen}
\author[20,21,22]{Licai Deng}
\author[20]{Kun Wang}
\author[23,24,25]{Lin Xiao} 
\author[26]{Lingjun Wang} 

\affil[1]{Physics Department and Tsinghua Center for Astrophysics (THCA), Tsinghua University, Beijing 100084, China}
\affil[2]{Beijing Planetarium, Beijing Academy of Science and Technology, Beijing 100044, China}
\affil[3]{Purple Mountain Observatory, Chinese Academy of Science, Nanjing 210008, China}
\affil[4]{Caltech Optical Observatories, California Institute of Technology, Pasadena, CA 91125, USA}
\affil[5]{Department of Particle Physics and Astrophysics, Weizmann Institute of Science, 234 Herzl St., 76100 Rehovot, Israel}
\affil[6]{Department of Astronomy, University of California, Berkeley, CA 94720-3411, USA}
\affil[7]{The Oskar Klein Centre \& Department of Astronomy, Stockholm University, AlbaNova, SE-106 91 Stockholm, Sweden}
\affil[8]{Department of Physics and Astronomy, Texas A\&M University, 4242 TAMU, College Station, TX 77843, USA}
\affil[9]{George P. and Cynthia Woods Mitchell Institute for Fundamental Physics \& Astronomy, USA}
\affil[10]{Division of Physics, Mathematics, and Astronomy, California Institute of Technology, Pasadena, CA 91125, USA}
\affil[11]{Departament de Física Quàntica i Astrofísica (FQA), Universitat de Barcelona (UB), c. Martí i Franquès, 1, 08028, Barcelona, Spain}
\affil[12]{Institut de Ciències del Cosmos (ICCUB), Universitat de Barcelona (UB), c. Martí i Franquès, 1, 08028, Barcelona, Spain} 
\affil[13]{Institut d'Estudis Espacials de Catalunya (IEEC), c. Gran Capità, 2-4, 08034, Barcelona, Spain}
\affil[14]{Ulugh Beg Astronomical Institute, Uzbekistan Academy of Sciences, Tashkent 100052, Uzbekistan}
\affil[15]{National University of Uzbekistan, Tashkent 100174, Uzbekistan}
\affil[16]{Yunnan Observatories, Chinese Academy of Sciences, Kunming 650216, China}
\affil[17]{Key Laboratory for the Structure and Evolution of Celestial Objects, Chinese Academy of Sciences, Kunming 650216, China}
\affil[18]{Department of Astronomy, Beijing Normal University, Beijing, People’s Republic of China}
\affil[19]{Center for Astronomical Mega-Science, Chinese Academy of Sciences, 20A Datun Road, Chaoyang District, Beijing, 100012, China}
\affil[20]{Department of Astronomy, China West Normal University, Nanchong, China}
\affil[21]{CAS Key Laboratory of Optical Astronomy, National Astronomical Observatories, Chinese Academy of Sciences, Beijing, China}
\affil[22]{School of Astronomy and Space Science, University of Chinese Academy of Sciences, Beijing, China}
\affil[23]{Department of Physics, College of Physical Sciences and Technology, Hebei University, Wusidong Road 180, Baoding 071002, China}
\affil[24]{Key Laboratory of High-precision Computation and Application of Quantum Field Theory of Hebei Province, Hebei University, Wusidong Road 180, 071002, Baoding, China}
\affil[25]{Research Center for Computational Physics of Hebei Province, Baoding, 071002, China}
\affil[26]{Astroparticle Physics, Institute of High Energy Physics, Chinese Academy of Sciences, Beijing 100049, China}
\affil[*]{wang\_xf@mail.tsinghua.edu.cn}
\begin{abstract}

\textbf{Superluminous supernovae are among the most energetic stellar explosions in the Universe, but their energy sources remain an open question. Here we present long-term observations of one of the closest examples of the hydrogen-poor subclass (SLSNe-I), SN~2017egm, revealing the most complicated known luminosity evolution of SLSNe-I. Three distinct post-peak bumps were recorded in its light curve collected at about $100$--350\,days after maximum brightness, challenging current popular power models such as magnetar, fallback accretion, and interaction between ejecta and a circumstellar shell. However, the complex light curve can be well modelled by successive interactions with multiple circumstellar shells with a total mass of about $6.8$--7.7\,M$_\odot$. In this scenario, large energy deposition from interaction-induced reverse shocks results in ionization of neutral oxygen in the supernova ejecta and hence a much lower nebular-phase line ratio of [O\,\textsc{i}] $\lambda6300$/([Ca\,\textsc{ii}] + [O\,\textsc{ii}]) $\lambda7300$ ($\sim 0.2$) compared with that derived for other superluminous and normal stripped-envelope SNe. The pre-existing multiple shells indicate that the progenitor of SN~2017egm experienced pulsational mass ejections triggered by pair instability within 2 years before explosion, in robust agreement with theoretical predictions for a pre-pulsation helium-core mass of 48--51\,M$_{\odot}$. Finally, this work shows that the final explosion product may be a black hole with about 40\,M$_{\odot}$, and has significant implication for the formation of such heavy black holes that have been recently observed by LIGO-Virgo gravitational wave detectors.}
  
\end{abstract}

\begin{document}
\maketitle

\section{Main}
At a distance of $128.8$\,Mpc (redshift $z=0.03063$), SN~2017egm (also known as Gaia17biu) is one of the closest known hydrogen-poor superluminous supernovae \cite{2011Natur.474..487Q, 2012Sci...337..927G} (SLSNe-I) found in the past decade. It was discovered \cite{2017TNSTR.591....1D} by the {\it Gaia} satellite on 23 May 2017 (UTC dates are used throughout this paper; JD = 2,457,897.40), at coordinates $\alpha = 10^{\rm hr}19^{\rm m}05.62^{\rm s}$, $\delta = +46^{\circ} 27' 14.08''$ (J2000). The supernova (SN) host, NGC~3191, is a barred Sbc galaxy with a stellar mass of a few $10^{10}\,{\rm M}_\odot$, much higher than in typical SLSNe-I \cite{2014ApJ...787..138L, 2015MNRAS.449..917L, 2016ApJ...830...13P}. The results of various metallicity diagnostics at the SN position are controversial, with $Z\sim0.6$\,Z$_\odot$ \cite{2018A&A...610A..11I} or $Z \approx 1.3$--2.6\,Z$_\odot$ \cite{2017ApJ...849L...4C, 2017ApJ...845L...8N, 2018ApJ...853...57B, 2018ApJ...858...91Y}.

Early-time ultraviolet (UV)-optical photometric and spectroscopic observations reveal the blue and luminous radiation from supernova ejecta at the hot photospheric phase \cite{2017ApJ...845L...8N, 2018ApJ...853...57B, 2018ApJ...858...91Y}. Magnetar-powered model provides a fit to the early-time multiband light curves, with the resultant model parameters being consistent with those of other well-observed SLSNe-I \cite{2017ApJ...845L...8N, 2017ApJ...850...55N}. However, as a steady magnetar wind model has difficulty in reproducing the triangle-shaped primary peak and late-time undulations in light curves, interaction between the ejecta and circumstellar material (CSM) is favored \cite{2017ApJ...851L..14W, 2018ApJ...853...57B, 2022ApJ...933...14H, 2022CoSka..52a..46T} despite lacking collision-induced narrow spectral lines \cite{2018ApJ...853...57B, 2018ApJ...858...91Y, 2019ApJ...871..102N}. Moreover, deep limits set on the radio emission from SN~2017egm disfavor an association with an on-axis gamma-ray burst jet driven by a powerful central engine \cite{2018ApJ...853...57B, 2018ApJ...856...56C}. 

The peculiar light curve and massive host galaxy of SN~2017egm, in contrast with most SLSNe-I, gained much attention and deserve deeper investigation. Here we present new UV-optical-near-infrared observations covering very late times (out to 380 rest-frame days after $V$-band maximum brightness; Methods), providing essential clues to the energy source of SN~2017egm and to final evolution of the progenitor star.


\subsection{Complex Light Curves}
\begin{figure}
\center
\includegraphics[angle=0,width=\textwidth] {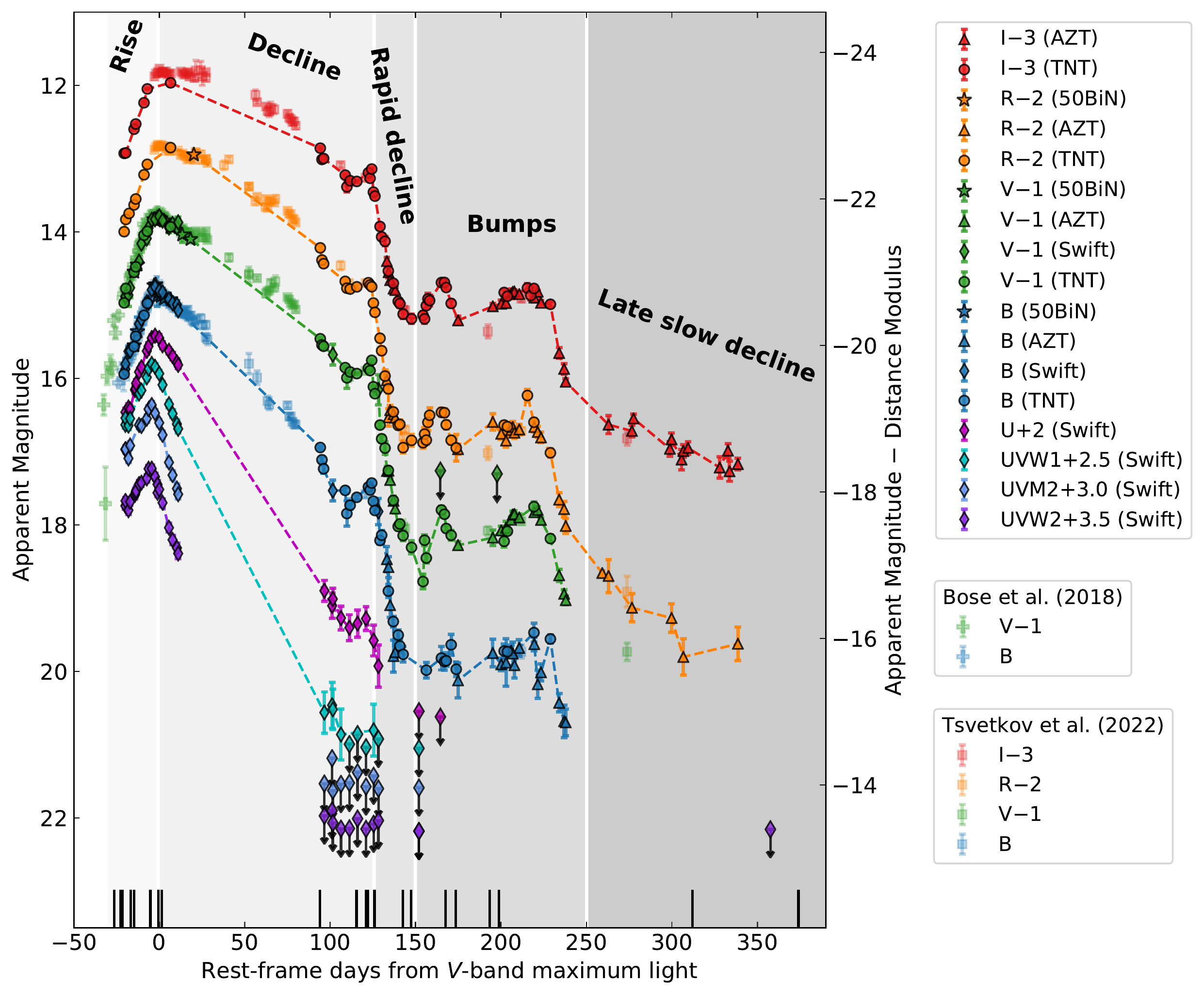}
\caption{{\bf{Multiband light curves of SN~2017egm}}. Our observations are obtained with the 50\,cm Binocular Network (50BiN; stars), the AZT-22 1.5\,m telescope (AZT; triangles), the {\it Swift}/Ultra-Violet/Optical Telescope ({\it Swift}, diamonds), and the Tsinghua-NAOC 0.80\,m telescope (TNT; circles). The $BVRI$ photometry from literature \cite{2018ApJ...853...57B, 2022CoSka..52a..46T} (plus and squares) is overplotted for comparison. The error bars show 1$\sigma$ uncertainties. For visual clarity, photometric data presented in this work are connected with dashed lines. The multiple phases (rise, decline, rapid decline, bumps, and late-time slow decline) of the light curves are individually shaded and labelled. The vertical black lines indicate the epochs when spectroscopic observations are available (Supplementary Table~\ref{Table: spec_Journal}). See Methods for details.}
\label{fig: Photometry}
\end{figure}

The UV-optical light curves of SN~2017egm (Fig.~\ref{fig: Photometry}) are characterized by five evolutionary phases, including quasi-linear rise ($t<0$\,d from $V$-band maximum light), first slow decline ($t \approx 0$--125\,d), rapid decline ($t \approx 125$--150\,d), bumps ($t \approx 150$--250\,d), and slow late-time decline ($t > 250$\,d). Similar evolution shared by multiband data indicates that the bumpy light curves are energized by a complex power source, rather than a result of drastic evolution of certain spectral lines. During $t=0-150$\,d, shorter-wavelength brightness declines faster, suggestive of rapid cooling in ejecta due to self-expansion and generally decreasing energy input. After that, no prominent decay is observed in $BVRI$ light curves for about 70 days, indicating the presence of extra power source at this period. The bolometric light curve of SN~2017egm can be obtained by fitting an UV-absorbed blackbody model \cite{2017ApJ...850...55N} to the spectral energy distribution (SED; see Methods for details). As seen in Fig.~\ref{fig: Lum_fit}, a triangle-shaped peak forms in the near-maximum light curve, in favor of an ejecta-CSM interaction power \cite{2017ApJ...851L..14W}. Right after the primary peak, SN~2017egm declines as fast as rapidly-declining events such as SN~2010gx; however, over a longer timescale (out to $t\approx 125$\,days), its flux dropped slowly in a manner similar to that of the slowly-evolving PTF12dam and SN 2018hti (Extended Data Fig.~\ref{fig: LTR_sne}). This suggests that extra energy is required to account for its relatively high luminosity at this phase. At later times, in contrast to the monotonic decay of PTF12dam, SN~2017egm exhibited a complex bolometric light curve. Its luminosity rises for days around $t \approx +120$\,days and decreased sharply about tenfold within $\sim 2$ weeks; it then rebrightened and displayed two bumps (one is spiked and the other is plateau-like) in the light curve at $t \approx 150$--250\,days. Both the triangle-shaped primary peak and the sharp drop following long-term slow decline are rarely seen in the growing sample of SLSNe-I with bumpy light curves \cite{2022ApJ...933...14H, 2023ApJ...943...42C}, but are reminiscent of a long-lived bright type II-P supernova iPTF14hls \cite{2017Natur.551..210A}, although the latter with much broader profile of bumps (Extended Data Fig.~\ref{fig: LTR_sne}; see Methods for more discussions).

\subsection{Model Fitting}
\label{subsec: model}
\begin{figure}
\center
\includegraphics[angle=0,width=0.5\textwidth]{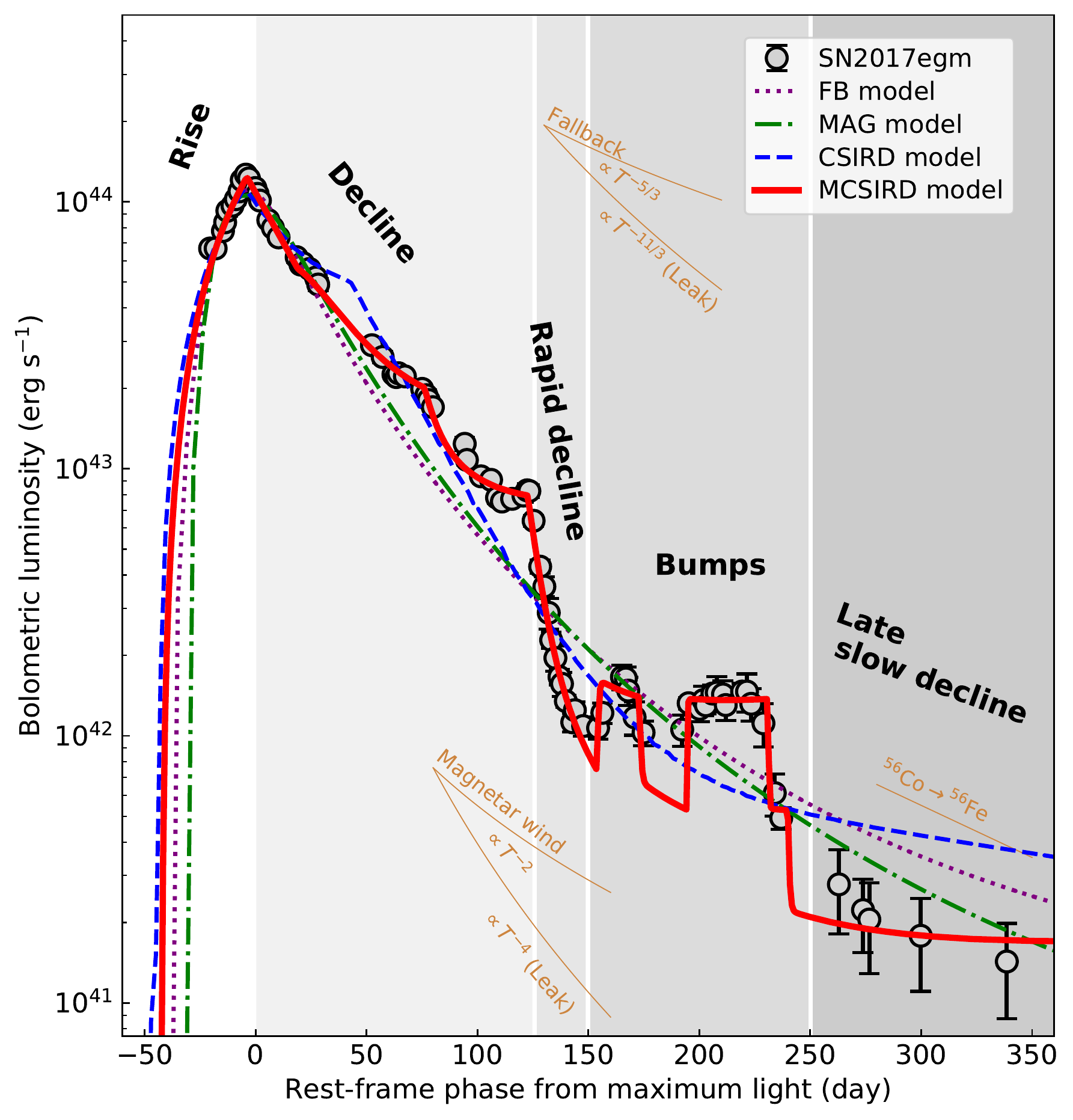}\includegraphics[angle=0,width=0.5\textwidth]{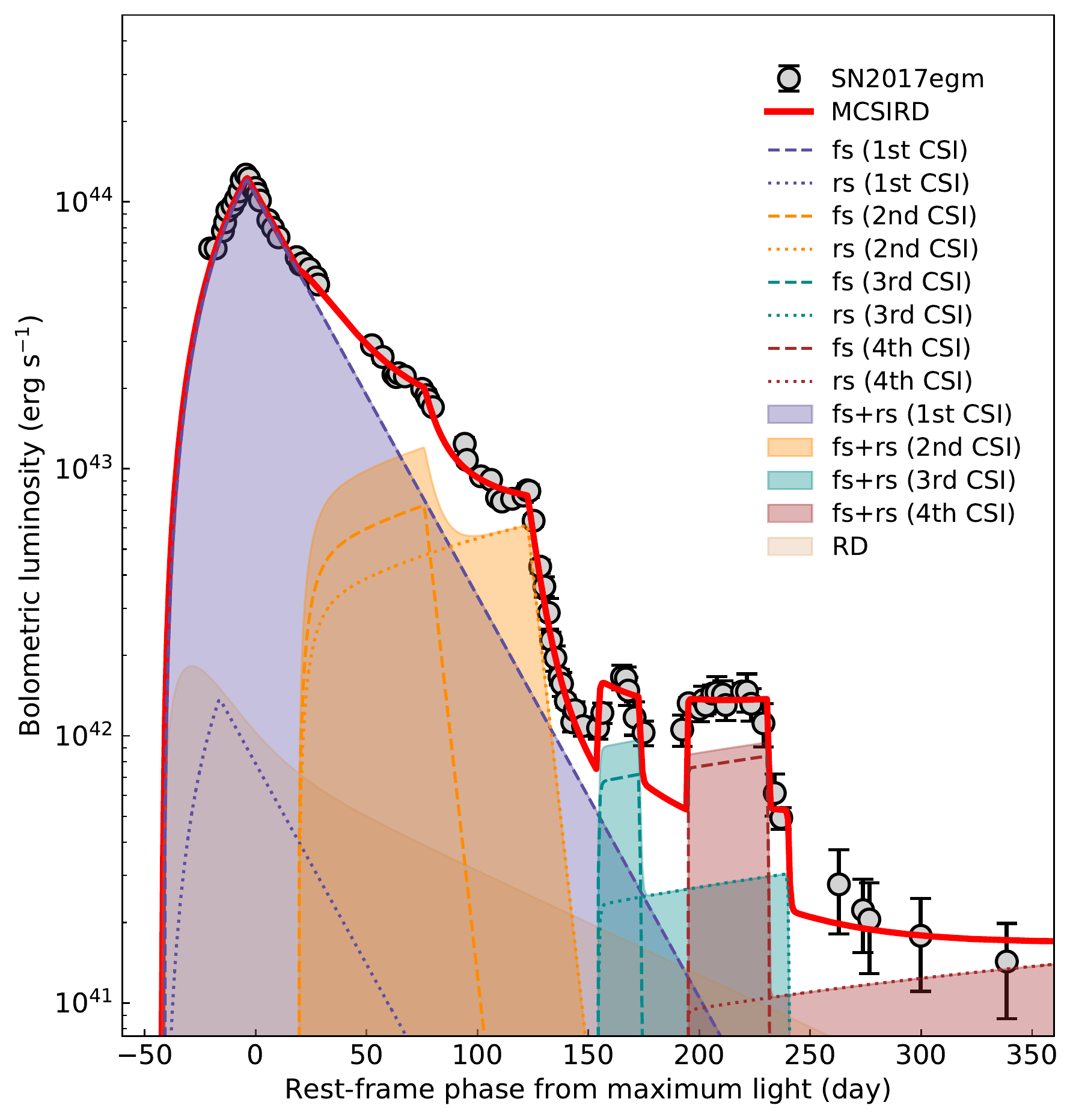}
\caption{{\bf{Model fits to the bolometric light curves of SN~2017egm}}. An MCSIRD model (red solid) can well explain the bolometric luminosity evolution of SN~2017egm (circles) by invoking four CSIs and $^{56}$Ni decay (RD) as the power mechanisms. The error bars show 1$\sigma$ uncertainties. {\it Left panel:} the MCSIRD model (red) is compared with fallback-powered \cite{2018ApJ...867..113M} (FB; purple dotted), magnetar-powered \cite{2017ApJ...850...55N} (MAG; green dotted-dashed), and single CSIRD-powered \cite{2012ApJ...746..121C} (CSIRD; blue dashed) models. The thin yellow solid lines (moved vertically for visual inspection) show the late-time input from the decay of $^{56}$Co to $^{56}$Fe, the fallback accretion ($\propto T^{-5/3}$ for no energy leakage and $\propto T^{-11/5}$ for full leakage), and the magnetar wind ($\propto T^{-2}$ for no energy leakage and $\propto T^{-4}$ for full leakage; here $T$ is the phase since explosion). The multiple phases of the light curve of this supernova (rise, decline, rapid decline, bumps, and late-time slow decline) of the light curves are individually grey-shaded and labelled. {\it Right panel:} for the MCSIRD model, the contributions from $^{56}$Ni decay (yellow) and each CSI (purple, orange, green, red) are shown with shaded regions. The radiations powered by forward shock (fs) and reverse shock (rs) are shown in dashed and dotted lines, respectively.}
\label{fig: Lum_fit}
\end{figure}

Post-peak bumps challenge the steady magnetar model and the continuous fallback accretion model (left panel of Fig.~\ref{fig: Lum_fit}). In particular, the rapid decay after $t \sim 125$ days is hardly reproduced with these two popular models, as the late-time magnetar wind (fallback accretion) power follows $P\propto T^{-2}$ ($P\propto T^{-5/3}$), or $P\propto T^{-4}$ ($P\propto T^{-11/3}$; here $T=t-t_\mathrm{exp}$ is the dynamical time after the explosion date 
$t_\mathrm{exp}$) if full energy leakage is considered. A variable magnetar wind model is proposed to explain a bumpy light curve, due to temporal changes of the wind thermalisation efficiency \cite{2022MNRAS.513.6210M} or fallback of stellar materials onto the central magnetar \cite{2018ApJ...857...95M, 2021ApJ...914L...2L}. In the fallback accretion model, episodic fallback should be invoked. However, for SN~2017egm with three post-peak bumps, either multiple fallback or frequent changes of thermalization efficiency are required, yet the mechanisms remain to be clarified.

Alternatively, the ejecta-CSM interaction can result in bumpy features; it is favored by the detection of broad H$\alpha$ emission in late-time spectra of a few SLSNe-I \cite{2015ApJ...814..108Y, 2017ApJ...848....6Y}. In the case of SN~2017egm, the interaction of SN ejecta with multiple CSM shells could provide a natural explanation for the sharp peak, the steep decline, and the late-time bumps. Moreover, the mid-infrared excess of the SED at $t \approx 100$--300\,d may indicate the existence of newly formed dust in the SN ejecta as a result of interaction, although the possibility of a light echo caused by ionization of pre-existing dust due to the supernova blast is not ruled out \cite{2022MNRAS.513.4057S}.

To account for the complicated light curve of SN~2017egm, i.e., the four bump features (including the primary peak), we adopt a hybrid power source model that involves multiple ejecta-CSM interaction (CSI) and radioactive decay (MCSIRD; Method). In this model, the expanding ejecta, with a mass of $2.55_{-0.28}^{+0.38}\,{\rm M}_\odot$ and a kinetic energy of $\sim 10^{51}$\,erg, collides successively with four CSM shells having masses of $4.09_{-0.25}^{+0.23}$\,M$_\odot$, $1.23_{-0.04}^{+0.05}$\,M$_\odot$, $0.85_{-0.07}^{+0.07}$\,M$_\odot$, and $1.1_{-0.1}^{+0.1}$\,M$_\odot$, respectively (Extended Data Table~\ref{tab: Lum_fit}). Here the MCSIRD model assumes that shock energy diffuses right behind the photosphere within the dense CSM \cite{2012ApJ...746..121C} rather than deep inside the ejecta \cite{2017ApJ...851L..14W}. As seen in the right panel of Fig.~\ref{fig: Lum_fit}, the SN emission around maximum light is powered mostly by the forward shock produced in the encounter between the ejecta and the nearest CSM shell. As the forward shock dominates over the reverse shock, the luminosity starts to decline several days after the forward shock breaks out of the optically thick part of CSM shell. This explains the formation of a sharp peak in the light curves. After $t \approx +19$\,days, the ejecta collides with the second CSM shell and contributes extra energy via shocks to the supernova radiation. The breakout time for the forward shock is $\sim 57$\,days, while for the reverse shock, it takes $103$\,days to pass through the ejecta. The termination of this reverse shock leads to the steep decay since $t=+125$~d. Interactions with the third and the fourth CSM shells are responsible for the two rebrightenings observed during $t=150-250$~d. The reverse shocks in both interactions are weaker than the forward shock, but persist longer. After the breakout of forward shock in fourth collision, the reverse shock and radioactive $^{56}$Co decay start to dominate the light-curve evolution ($t \approx 250$--350\,days). The tail luminosity can place a constraint on the mass of $^{56}$Ni synthesized in the explosion, which turns out to be 0.07\,M$_\odot$.

\subsection{Spectral Signatures for power source}
\label{subsec: Inter_sign}

As the complex, bumpy light curve of SN~2017egm strongly implies multiple collisions between the expanding ejecta and the surrounding CSM shells, we examine our spectra series (spanning $\sim 400$\,days; Extended Data Fig.~\ref{fig: spec}) to search for possible narrow lines that arise from the unshocked slowly-moving CSM shell. As seen in Extended Data Figs.~\ref{fig: NaI_HeI} and \ref{fig: galaxy}, both narrow H\,\textsc{i} and He\,\textsc{i}$~\lambda$5876 lines can be detected in the supernova spectra, and the line profile of latter seems to show some variations at different phases, with a velocity width varying from 200 to 500 km s$^{-1}$ during $t=+141-374$~d. However, similar H/He features are also visible in the host-galaxy spectrum of SN 2017egm, indicating that the detection of such features in supernova spectra may suffer from host-galaxy contamination. In addition, the spectra of this supernova lack the narrow carbon and oxygen lines recently observed in the interaction-powered type~Icn supernovae \cite{2022Natur.601..201G, 2022ApJ...927..180P}. However, it has been pointed out that the excitation of helium, carbon, and oxygen in CSM shells depends on the distribution, motion, and ionization of the CSM \cite{2013ApJ...773...76C}. Moreover, such lines have been previously shown \cite{2014ApJ...785...37B} to be transient and difficult to detect, so they could have been missed in our spectroscopic series. Nevertheless, a broader emission feature coincident with He\,\textsc{i} $\lambda$10,830, with a velocity width of about 6300 km s$^{-1}$, is detected in the $t \approx 193$\,d near-infrared spectrum (Extended Data Fig.~\ref{fig: comp_NIR}), which may arise from ejecta or partly from fast-moving circumstellar helium that has been accelerated by previous ejecta-CSM interaction (Method).

\begin{figure}
\center
\includegraphics[angle=0,width=1\textwidth]{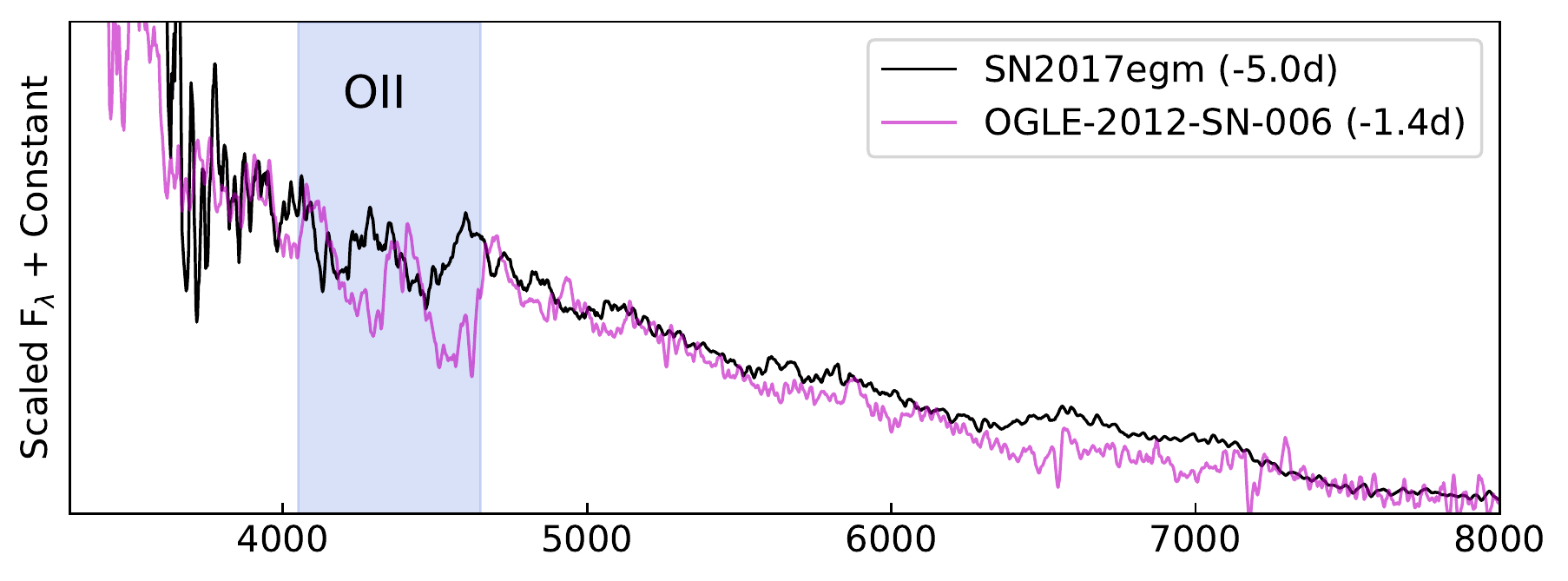}
\includegraphics[angle=0,width=1\textwidth]{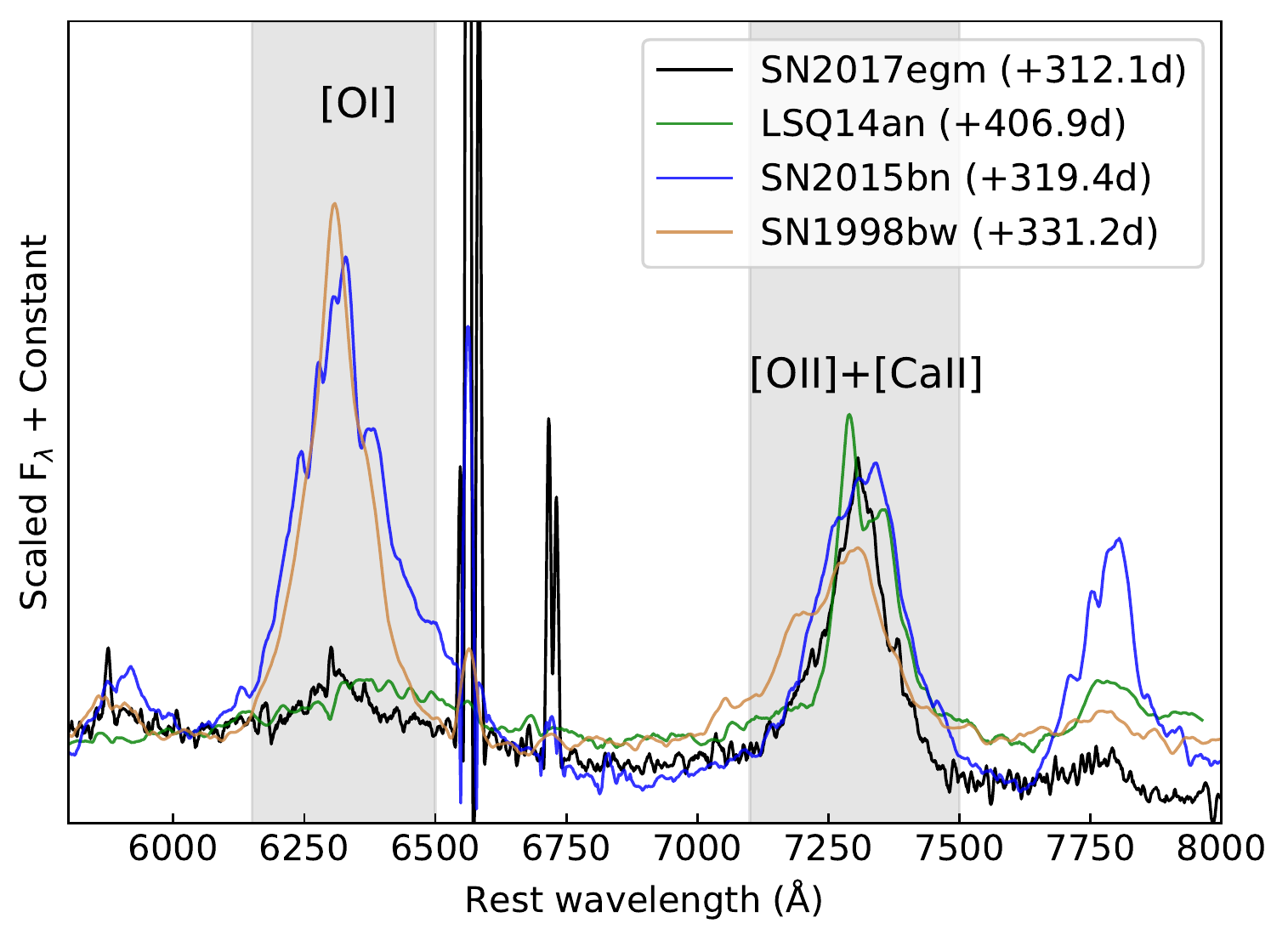}
\caption{{\bf{Spectral signs of powerful energy deposition for SN~2017egm.}} SN~2017egm shows early-time O\,\textsc{ii} absorption ({\it upper panel}) and late-time low line-intensity ratio of [O\,\textsc{i}]/([Ca\,\textsc{ii}] + [O\,\textsc{ii}] ({\it lower panel}). The comparison SNe include type Ibn supernova OGLE-2012-SN-006 (purple) \cite{2015MNRAS.449.1941P}, type Ic-BL supernova SN~1998bw (yellow) \cite{2001ApJ...555..900P} and SLSNe-I (LSQ14an, green; SN~2015bn, blue) \cite{2017ApJ...835...13J}. All of the spectra were smoothed with a Savitzky-Golay filter.}
\label{fig: comp_spec}
\end{figure}

One may question whether other spectral features of SN~2017egm fit with the CSI scenario. In that context, we notice that the early-time O\,\textsc{ii} complex at 3500--5000\,\AA\ is not unique to the SLSN-I population; it was also detected in an type Ibn supenova (OGLE-2012-SN-006 \cite{2015MNRAS.449.1941P}; upper panel of Fig.~\ref{fig: comp_spec}). In the magnetar-powered model, the appearance of O\,\textsc{ii} features can be ascribed to nonthermal excitation by a magnetar \cite{2016MNRAS.458.3455M}, while detections of both O\,\textsc{ii} and narrow emission lines of H\,\textsc{i} and He\,\textsc{i} in OGLE-2012-SN-006 suggest that the O\,\textsc{ii} absorption features could also be related to an interaction process. 

\begin{figure}
\center
\includegraphics[angle=0,width=1\textwidth]{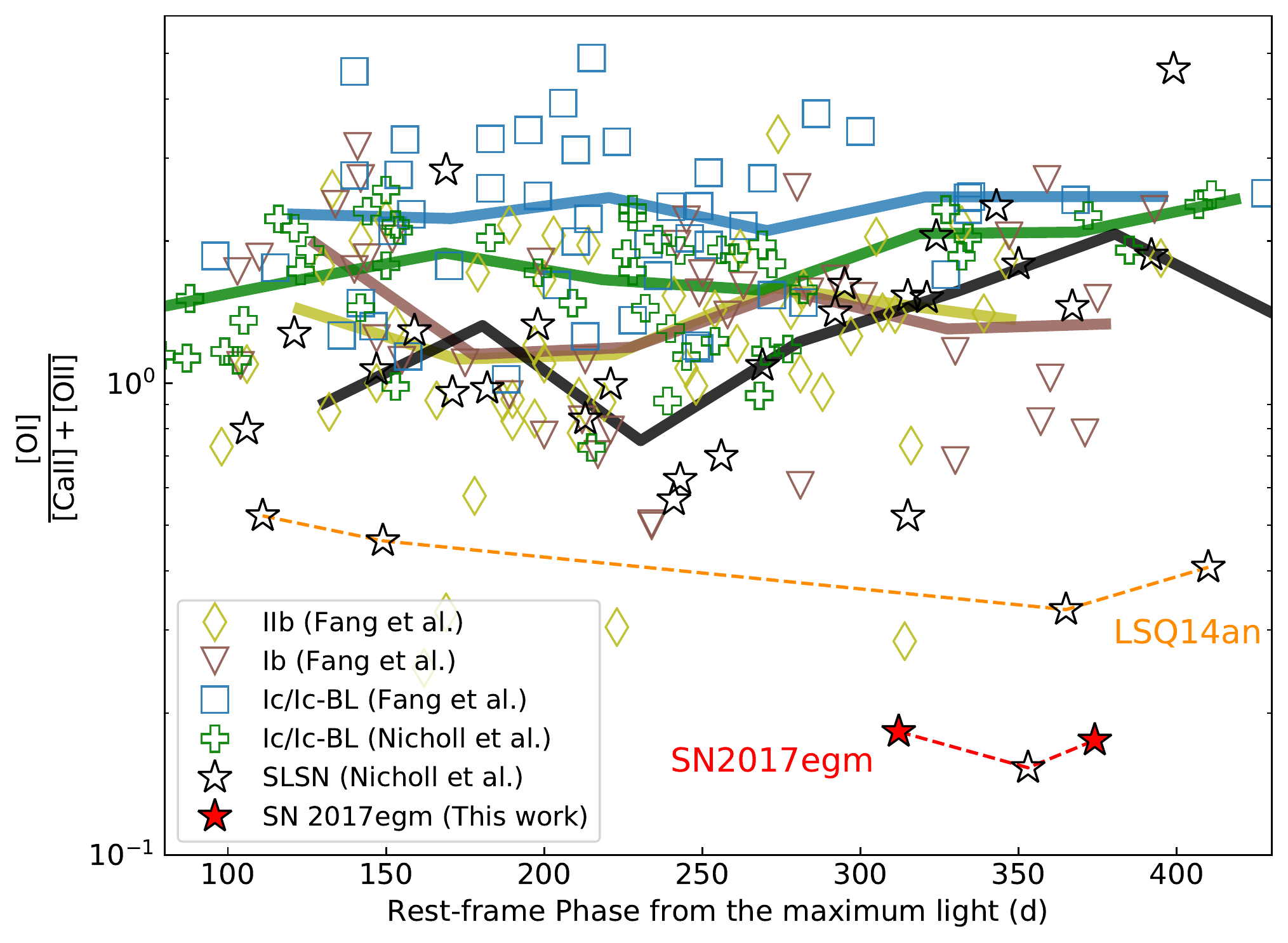}
\caption{{\bf{The nebular-phase [O\textsc{i}]/([Ca\textsc{ii}] + [O\textsc{ii}]) ratios of SLSNe-I and other types of envelope-stripped supernovae.}} The measured flux ratios of SN~2017egm (red filled stars, this work) are compared with previous sample of SLSNe-I (empty stars), type IIb supernovae (yellow diamonds), type Ib supernovae (brown triangles) and type Ic/Ic-BL supernovae (blue squares and green pluses) \cite{2018ApJ...864...47F,2019NatAs...3..434F, 2022ApJ...928..151F, 2019ApJ...871..102N}. The LSQ14an and SN~2017egm data are connected with orange and red dashed lines, respectively. For each comparison sample, the corresponding solid line represents the average line ratios for bins of 50 or 100\,days (each supernova contributes only one data point or one average value per bin). We caution that the [O\textsc{i}] luminosity of SN~2017egm at $t \lesssim 250$\,d could be overestimated owing to blends of multiple features around 6300\,\AA\, (Extended Data Fig.~\ref{fig: spec_OCa}, top left). Thus, the ratios measured during that period from Ref.~\cite{2019ApJ...871..102N} are not shown here. }
\label{fig: comp_OI_CaII}
\end{figure}

Late-time optical spectra of SN~2017egm display distinct properties, showing an unusually low flux ratio of [O\,\textsc{i}] $\lambda$6300/([Ca\,\textsc{ii}] + [O\,\textsc{ii}]) $\lambda$7300 (i.e., $0.15\pm0.01$ at t=+353\,d \cite{2019ApJ...871..102N}; $0.18\pm0.01$ at t=+312\,d and $0.17\pm0.01$ at t=+374\,d, this work; lower panel of Fig.~\ref{fig: comp_spec} and Extended Data Fig.~\ref{fig: spec_OCa}). The values of these ratios are much smaller than those of other superluminous and normal-luminosity stripped-envelope supernovae \cite{2018ApJ...864...47F,2019NatAs...3..434F,2019ApJ...871..102N, 2022ApJ...928..151F}, as shown in Fig.~\ref{fig: comp_OI_CaII}. Such a low line ratio might suggest that the progenitor of SN~2017egm is much smaller than the typical stars that end as core-collapse supernovae. This is because in the context of ordinary stripped-envelope supernovae at nebular phases (most oxygen is neutral), the line ratio of [O\,\textsc{i}]/[Ca\,\textsc{ii}] is argued to be positively correlated with the CO core mass of the progenitor and hence with the zero-age main-sequence mass \cite{2022ApJ...928..151F}. However, an extremely low-mass star is not expected to produce an SLSN powered by a magnetar, which are usually born as the explosions of rapidly rotating massive stars \cite{2015Natur.528..376M}, or to expel large amounts of H-poor material before explosion so as to produce the massive CSM shells required by the interaction model (i.e., the MCSIRD model) for SN~2017egm. Moreover, an extremely low-mass core or progenitor is not favored by the light-curve modelling, as the inferred ejecta masses (2--4~M$_\odot$ \cite{2017ApJ...845L...8N}; $\sim30$~M$_\odot$ \cite{2017ApJ...851L..14W}; 2--3~M$_\odot$, this work) are comparable to or much larger than the typical ejecta mass range for normal stripped-envelope SNe (1--5\,M$_\odot$ \cite{2011ApJ...741...97D}).

Thus, an alternative and more likely explanation for the low ratio of $\lambda6300/\lambda$7300 is that neutral oxygen in the ejecta is ionized and that ionized oxygen features dominate in the late-time spectra \cite{2019ApJ...871..102N}. From MCSIRD model fits, the inferred temperatures at both the forward and reverse shock fronts in all four collisions are always above several $10^7$~K (Methods),  higher than the temperature required for ionization of oxygen ($1.58\times10^{5}$\,K, corresponding to the ionization energy of 13.618\,eV). Unlike the central-engine scenario where energy is initially released and reprocessed in the inner region of the ejecta, interaction-induced shocks form at the ejecta-CSM interface and reverse shock inward would release amounts of power in the outer layer of the ejecta, which more easily leads to the outwards extension of the central ionized region. Hence, a strong reverse shock is necessary for reduction of the neutral oxygen abundance of the ejecta in the case of SN~2017egm, where the efficient energy deposition rate is inferred from MCSIRD model to be $\sim10^{41}$~erg~s$^{-1}$ during $t=300-400$~d. Based on the nebular spectral modelling with high input power of $10^{41-43}$~erg~s$^{-1}$ at this phase \cite{2017ApJ...835...13J, 2019A&A...621A.141D}, ionization of oxygen depends on the properties of both the ejecta and the power source; more specifically, lower mass, lower clumping and larger energy deposition for the ejecta can result in higher ionization and hence lower neutral oxygen abundance. This is consistent with the observation of low flux ratio of $\lambda6300/\lambda7300$ in SN~2017egm, as the inferred ejecta mass ($M_\mathrm{ej}\approx2.55$\,M$_\odot$) is comparable to the lowest mass ($M_\mathrm{ej} = 3$\,M$_\odot$) adopted in those spectral model simulations. Note that the $\lambda6300$ feature is dominated by [O~\textsc{i}] in this supernova (Extended Data Fig.~6) although it could be affected by [Fe~\textsc{ii}] in some spectral models \cite{2019A&A...621A.141D}. For heavier ejecta (for example, $\gtrsim9$\,M$_\odot$ for LSQ14an), reduction of the neutral oxygen abundance requires lower clumping and/or more energy deposition \cite{2017ApJ...835...13J, 2019A&A...621A.141D}.

\subsection{Discussion}
 \label{sec: discussion}

Based on the properties of CSM inferred from the MCSIRD model fit, the progenitor of SN~2017egm probably underwent at least four violent eruptions and ejected a total mass ($M_\mathrm{CSM,tot}$) of 6.8--7.7\,M$_\odot$, within $T_\mathrm{w}\sim2/(v_\mathrm{w}/1000$\,km\,s$^{-1}$) years before its final explosion ($v_\mathrm{w}$ is the CSM velocity; Methods). For a typical velocity range from 10 to several 1000\,km\,s$^{-1}$, the mass-loss rate ($\dot{M}=M_\mathrm{CSM,tot}/T_\mathrm{w}$) inferred for the progenitor star at this eruptive period is about 0.03--10\,M$_\odot$\,yr$^{-1}$. This is in contradiction with the results (i.e., $\dot{M}<10^{-3}$\,M$_\odot$\,yr$^{-1}$) from deep radio upper limits based on the synchrotron self-absorption (SSA) model \cite{2018ApJ...856...56C}. However, the SSA model and its prediction might be not applicable in the case of such a high mass-loss rate, as the well-known interacting type IIn supernova SN 2010jl, with a high mass-loss rate of 0.1\,M$_\odot$\,yr$^{-1}$ inferred from its X-ray light curve, was also found to show weak radio emission \cite{2015ApJ...810...32C} that is below the upper limit for SN~2017egm. Such mass ejection is always more intense than binary Roche-lobe overflow ($\dot{M}<0.1$\,M$_\odot$\,yr$^{-1}$) and typical stellar wind of evolved massive stars ($\dot{M}<10^{-3}$\,M$_\odot$\,yr$^{-1}$), but is consistent with giant eruptions of luminous blue variable (LBVs; $\dot{M}<10$\,M$_\odot$\,yr$^{-1}$ with typical velocity of several 100 km s$^{-1}$) \cite{2017hsn..book..403S}. However, the mechanism of triggering frequent LBV eruptions within years is unknown. Moreover, LBV eruption is usually related to hydrogen ejection, which might be inconsistent with the hydrogen-free CSM shells for this SN. 

\begin{figure}
\center
\includegraphics[angle=0,width=1\textwidth]{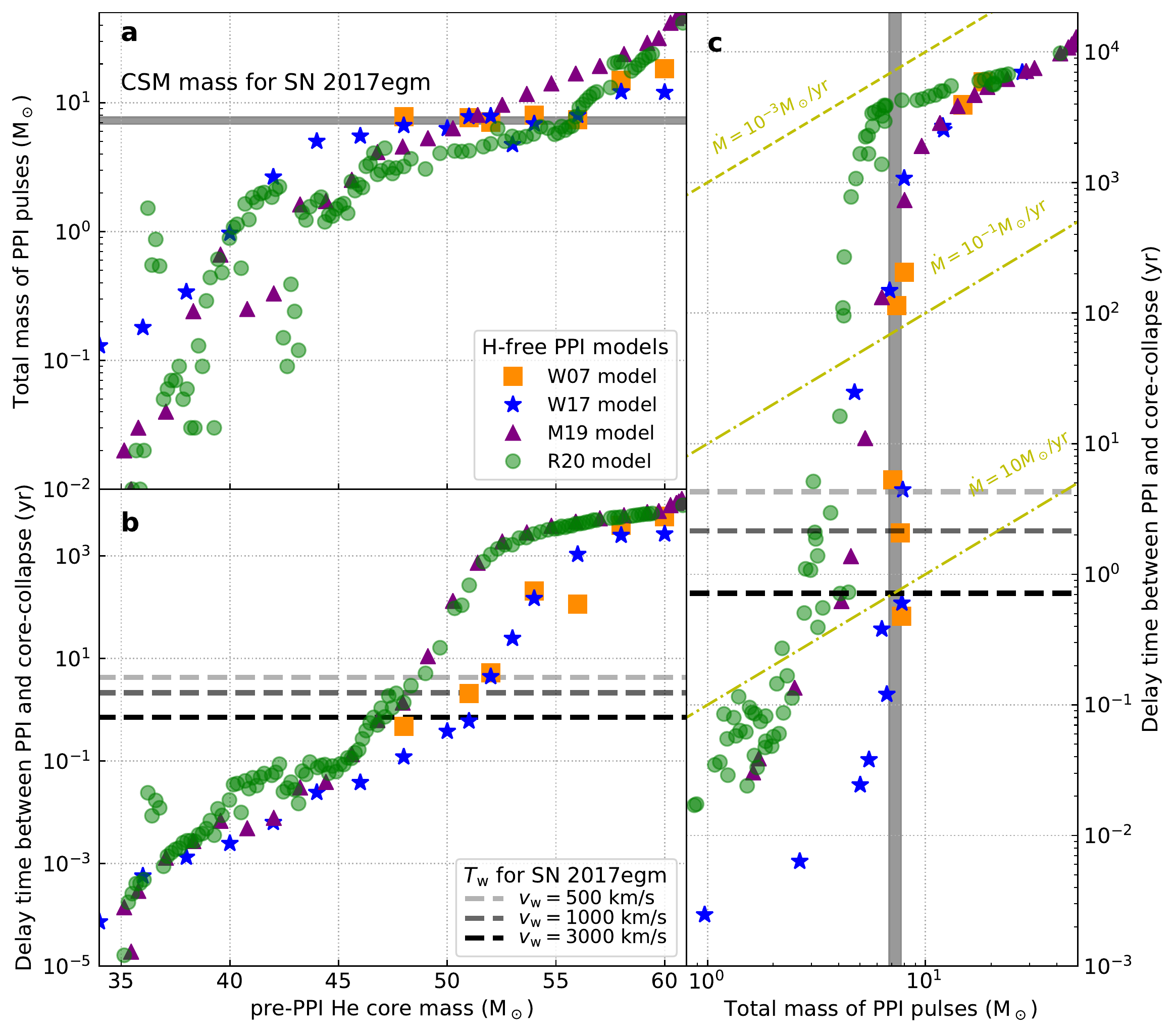}
\caption{{\bf{Comparison with hydrogen-free PPI models.}} The inferred properties of SN~2017egm are shown against four hydrogen-free PPI models of W07\cite{2007Natur.450..390W} (orange squares) , W17 \cite{2017ApJ...836..244W} (blue stars), M19 \cite{2019ApJ...882...36M} (purple triangles), and R20 \cite{2020A&A...640A..56R} (green circles), in the diagrams of PPI-driven mass loss versus pre-pulsation helium core mass ({\bf{a}}), delay time between the PPI onset and core-collaspe explosion versus pre-pulsation helium-core mass  ({\bf{b}}) and delay time versus PPI-driven mass loss ({\bf{c}}). The narrow grey shade shows the total mass range (within 1$\sigma$ uncertainty) of four CSM shells inferred from light-curve modeling of SN~2017egm. Assuming velocities of $v_w=$ 500, 1000, 3000 km s$^{-1}$, the inferred delay times between core collapse and the eruption related to the outermost shell ($T_\mathrm{w}$) are displayed with gray dashed lines. The yellow lines represent the steady mass ejections with mass-loss rates of $\dot{M}=10^{-3}, 0.1, 10$~M$_\odot$ yr $^{-1}$.}
\label{fig: PPI_mass}
\end{figure}

Instead, the pulsational pair-instability (PPI) mechanism provides an attractive alternative that naturally explains the remarkable mass loss history of this event. In this scenario, a very massive star undergoes episodic mass ejections due to non-terminal bursts of thermonuclear oxygen burning triggered by pair instability, until the core is no longer unstable, and eventually evolves to core collapse \cite{2007Natur.450..390W, 2017ApJ...836..244W}. Stellar rotation is required for formation of a millisecond magnetar or a black hole + disk system after core collapse, which is essential to powering a successful explosion. The PPI-driven mass ejection from fast rotating progenitors can be deficient in hydrogen but rich in helium and oxygen \cite{2012ApJ...760..154C}, in agreement with the lack of H features in the spectra of SN~2017egm. As shown in Fig.~\ref{fig: PPI_mass}, the inferred mass of four CSM shells (6.8--7.7 M$_\odot$) and initial time of eruptions (i.e., 0.5--2 years before explosion for a typical velocity of PPI-triggered pulse of $v_\mathrm{w}=1000-4000$ km s$^{-1}$) are roughly consistent with the hydrogen-free PPI models for pre-pulsation helium-core mass of $48$\,M$_\odot$ (W07 model \cite{2007Natur.450..390W}) and 51\,M$_\odot$ (W07 \cite{2007Natur.450..390W} and W17 \cite{2017ApJ...836..244W} models), which predict 4--6 PPI-driven eruptions with a total mass of 7.7--7.8\,M$_\odot$. Note that both mergers of PPI-driven pulses and collisions with pre-PPI mass loss can decelerate shells and affect the final CSM configuration around the exploding star. Given the inferred supernova ejecta mass of 2.55 M$_\odot$ from MCSIRD model, the core-collapse explosion may have resulted in a black hole with a mass of about 37--41 M$_\odot$. Thus, this work shows that black holes with tens of solar masses, which have been recently detected by LIGO-Virgo gravitational wave detectors \cite{2019PhRvX...9c1040A}, can be produced through PPI + core-collapse channel, and not exclusively via the merger of lighter black holes. We caution that the predictions for PPI pulses are sensitive to the assumptions (e.g., metallicity, hydrodynamics, stellar evolution and pulse definition) adopted in PPI models. The hydrogen-free PPI models of M19 \cite{2019ApJ...882...36M} and R20 \cite{2020A&A...640A..56R} require a much longer pulsation period (from hundreds to thousands of years) to produce such massive CSM (6.8--7.7 M$_\odot$). 

A possible challenge to the PPI scenario is that the high environmental metallicity (1.2--2.7~Z$_\odot$ favored by most of our measurements based on various host-line diagnostics; Methods and Extended Data Fig.~\ref{fig: metallicity}) inferred for SN~2017egm is higher than the robust upper limit of $\sim 1/3$ Z$_\odot$ \cite{2017ApJ...836..244W} (or $\sim 1/2$ Z$_\odot$ \cite{2019ApJ...887...72L}) beyond which the progenitor may not retain sufficiently massive helium core to trigger PPI. If SN~2017egm has a single stellar origin, the PPI scenario requires (1) a metal-rich progenitor with greatly reduced mass-loss rate before oxygen burning stage or (2) a metal-poor star that somehow exploded in the outskirt of a metal-rich host galaxy. However, the progenitor may initially have been a massive binary; a merger of two massive stars could also result in a heavy pair-unstable core \cite{2019ApJ...876L..29V}. 

Thus, SN~2017egm provides new insights into the evolution of massive stars and perhaps the formation of massive helium cores in high-metallicity environments.

\clearpage


\section{Methods}
\subsection{Photometry}

We commenced optical photometry of SN~2017egm with several instruments, including the 50\,cm Binocular Network (50BiN) \cite{2013IAUS..288..318D} at the Delingha Station of the Purple Mountain Observatory in China, the AZT-22 1.5\,m telescope (AZT) at Maidanak Astronomical Observatory \cite{2018NatAs...2..349E} in Uzbekistan, and the Tsinghua-National Astronomical Observatories of the Chinese Academy of Sciences (NAOC) 0.80\,m telescope (TNT) \cite{2008ApJ...675..626W, 2012RAA....12.1585H} at Xinglong Observatory. All of the data were reduced with ZrutyPhot pipeline (J. Mo et al., manuscript in preparation). Data reduction and point-spread-function (PSF) fitting methods of ZrutyPhot follow the standard Image Reduction and Analysis Facility (\texttt{IRAF}) routines, including corrections for bias, flat field and removal of cosmic rays in the charge-coupled device images. To remove host-galaxy contamination from the TNT and AZT data, which is especially important for late-time photometry, we perform template subtraction with \texttt{HOTPANTS} technique \cite{2015ascl.soft04004B} using multiband templates obtained on 21 December 2020 (for TNT) and 2 April 2021 (for AZT), respectively. Finally, the instrumental magnitudes measured through the PSF fitting method in \texttt{IRAF} were calibrated based on ten local stars (Supplementary Table~\ref{Table: referstar}) from the Panoramic Survey Telescope \& Rapid Response System (Pan-STARRS) catalogue \cite{2012ApJ...750...99T, 2020ApJS..251....6M}. Here the Pan-STARRS magnitudes are converted to the $BVRI$ photometric system based on previously established transformation functions\cite{2012ApJ...750...99T}. The final flux-calibrated $BVRI$ photometry is presented in Supplementary Table~\ref{Table: phot}. The magnitude errors include the Poisson noise of photons, the flux-calibration errors of standard stars and the errors of colour-term corrections.

The observations with the Ultra-Violet/Optical Telescope (UVOT) \cite{2005SSRv..120...95R} onboard the Neil Gehrels {\it Swift} Observatory mostly cover near-maximum phase and post-peak 100--200\,days. We obtained the photometry using the data-reduction pipeline (including template subtraction) of the Swift Optical/Ultraviolet Supernova Archive \cite{2014Ap&SS.354...89B}. The template images used for subtraction were obtained on 2018 Jul. 6th (corresponding to t$\sim$ 370 d after the $V$-band maximum) when the SN was very faint. The source counts were yielded by subtracting the background counts measured using the template image with a $5''$ aperture. After calibration with updated zeropoints for the UVOT filters \cite{2011AIPC.1358..373B}, the resultant UVOT photometry is listed in Supplementary Table~\ref{Table: phot_swift}.

A commonly used method of determining the peak epoch is to fit near-maximum light curve with a low-order polynomial function. Note, however, that the near-maximum V-band light curve exhibits a sharp peak formed by a quasi-linear rise and decline, which may not be well reproduced by a polynomial function. Thus, the observed epoch (modified Julian date (MJD) 57926.72) of the brightest V-band observation is adopted as the maximum-light epoch in our analysis.

Fig.~\ref{fig: Photometry} shows the multiband light curves of SN~2017egm from about $-20$\,days to +338\,days from $V$-band maximum brightness. The early-time data obtained with TNT and 50BiN are roughly consistent with the photometry obtained by other observational campaigns \cite{2018ApJ...853...57B, 2017ApJ...845L...8N}. The AZT and TNT datasets collected after MJD $58023$ show complicated multipeak evolution in the $BVRI$ light curves, further complementing the late-time dataset of SN~2017egm \cite{2022ApJ...933...14H, 2022CoSka..52a..46T} and greatly expanding our recognition of the long-term evolution of SLSNe-I. 

The bolometric luminosity evolution and corresponding errors of SN~2017egm were estimated by using a least square method (\texttt{scipy.curve\_fit} \cite{2020SciPy-NMeth}) to fit the SED with a UV-absorbed blackbody model \cite{2017ApJ...850...55N}. The SED was constructed from the UV-optical photometry reported in this work (Supplementary Tables \ref{Table: phot} and \ref{Table: phot_swift}) as well as the previously published $rizJHK$-band data \cite{2018ApJ...853...57B} and $BVRI$-band data \cite{2022CoSka..52a..46T}. After $t \approx +274$\,d, only $RI$-band photometry is available. Given that the temperature of SLSNe-I usually tends to remain unchanged in nebular phase \cite{2013ApJ...770..128I}, late-time $BV$-band photometry was estimated by assuming constant $B-R$ and $V-R$ colours, and it was further combined with $RI$ data to yield the bolometric luminosity at that phase. Before the fits, data were corrected for the Galactic extinction of $E(B-V)=0.0097$\,mag \cite{2011ApJ...737..103S} and the cosmological redshift of $z=0.03063$ \cite{2017ApJS..233...25A}. The distance modulus is derived as 35.55\,mag assuming a Lambda-cold-dark-matter ($\Lambda$CDM) cosmological model, with matter density parameter $\Omega_{M} = 0.27$, dark energy density parameter $\Omega_{\Lambda} = 0.73$, and Hubble constant H$_{0}$ = 73\,km\,s$^{-1}$\,Mpc$^{-1}$.
\\

\subsection{Multiple Ejecta-CSM Interaction and Radioactive Decay (MCSIRD) model}

We combine a semi-analytical hybrid model (single ejecta-CSM interaction and radioactive decay; CSIRD) \cite{2012ApJ...746..121C} and a multiple interaction model (developed based on CSIRD model) \cite{2018ApJ...856...59L} into a complex power-source model involving multiple interactions and radioactive decay (MCSIRD). Assume that a portion of the stellar matter (mass $M_\mathrm{ej}$) ejected in core collapse of a massive star encounter $N$ shells of pre-existing dense CSM (with mass $M_{\mathrm{CSM},k}$, inner radius $R_{\mathrm{CSM,in},k}$ and inner-radius density $\rho_{\mathrm{CSM,in},k}$; here $k$ is the sequential number of CSM shells). In a similar way to the CSIRD formula, the MCSIRD-powered supernova luminosity can be yielded as
\begin{equation}
L(t)= \frac{1}{t_\mathrm{diff,RD}}\int^{T}_{0}e^{\frac{t'-T}{t_\mathrm{diff,RD}}}P_\mathrm{RD}(t')\mathrm{d}t'+\sum^{N}_{k}\frac{1}{t_{\mathrm{diff,CSI},k}}\int^{T}_{t_{\mathrm{i},k}-t_\mathrm{exp}}e^{\frac{t'-T}{t_{\mathrm{diff,CSI},k}}}\left[P_{\mathrm{fs},k}(t')+P_{\mathrm{rs},k}(t')\right]\mathrm{d}t'\, ,
\label{eq: mcsird}
\end{equation}
where $T=t-t_\mathrm{exp}$ is dynamical time after explosion date ($t_\mathrm{exp}$), $t_{\mathrm{i},k}$ is the initial time of $k$-th interaction, $P_\mathrm{RD}$, $P_{\mathrm{fs},k}$ and $P_{\mathrm{rs},k}$ are the input powers from $^{56}$Ni+$^{56}$Co decay, forward shock and reverse shock, $t_\mathrm{diff,RD}\sim\kappa (M_{\mathrm{ej}}+\sum^{N-1}_k M_{\mathrm{CSM},k}+M_{\mathrm{CSM,th},N})/(13.8 c r_{\mathrm{ph},N})$ and $t_{\mathrm{diff,CSI},k}\sim \sum^N_k \kappa M_{\mathrm{CSM,th},k}/(13.8 c r_{\mathrm{ph},k})$ are the diffusion timescales for the radioactivity power and the radiative shock power produced during the interaction with $k$-th CSM shell, respectively. Here $c$ is the speed of light, a $M_{\mathrm{CSM,th},k}$ is the mass of optically thick part of the ejecta located behind the photospheric radius ($r_{\mathrm{ph},k}$). In this model, the initial times of interactions ($t_{\mathrm{i},k>1}$) serve as free parameters, and can be used to estimate inner radius of corresponding CSM shell as $R_{\mathrm{CSM,in},k}=R_{\mathrm{CSM,in},k-1}+(t_{\mathrm{i},k}-t_{\mathrm{i},k-1})v_{\mathrm{ej},k}$ \cite{2018ApJ...856...59L}, where $v_{\mathrm{ej},k}$ is the velocity of the ejecta at that phase.

Collisions of CSM shells are ignored in this model. They are often expected to occur in PPI scenario and can result in bright transients and merger-induced complex configuration of CSM. However, our semi-analytical model has difficulty including shell collisions, which requires a detailed numerical simulation for the stellar evolution of massive stars. Note that the starting times of collisions are in a wide range, varying mostly from decades before core collapse to thousands of years after that \cite{2007Natur.450..390W, 2017ApJ...836..244W}. Thus, collisions may not need to be always considered for a relatively short period of supernova light curve (with a phase of $\sim380$ days).

The MCSIRD model as well as two proposed interaction models \cite{2012ApJ...746..121C, 2018ApJ...856...59L} are built based on assumption of self-similar evolution for 
an interaction region \cite{1982ApJ...258..790C}, which requires a power-law density profile for both supernova ejecta and CSM shell. The ejecta configuration is assumed to be $\rho\propto r^{-n}$ for the outer part and $\rho\propto r^{\delta}$ for the inner part. However, fast-expanding ejecta could evolve into more complicated configurations after colliding with CSM shell, as a sharp cold dense layer might form between forward and reverse shocks. A high value of power-law index $n$ might be able to describe the sharp outer boundary of density profile of post-collision ejecta, whereas numerical simulation is necessary to give more accurate model predictions for properties of the whole interaction region. As both explosion mechanism for a PPI+core-collapse channel and effect of pre-PPI wind on final CSM configuration are not well understood, the semi-analytical MCSIRD model can provide a guide to explore the parameter space.

\subsection{MCSIRD Model fitting}
Motivated by the four bumps seen in the light curve of SN~2017egm, we perform a fit with the MCSIRD model involving interactions with four CSM shells. This model requires spherically symmetric structure for both ejecta and CSM. It is reasonable for SN~2017egm, as spectropolarimetric observations of this supernova reveal a null polarization result \cite{2019MNRAS.482.4057M} or an intrinsic polarization at $\sim0.2$\%$-0.8$\% level \cite{2018ApJ...853...57B, 2020ApJ...894..154S}, indicating a low or modest asphericity in the configuration of ejecta.

Here we assume an ejecta configuration with $\delta=0$ and $n=12$ being the power-law exponents for the inner and outer density profiles, as well as a constant-density profile for each CSM shell which is more appropriate than a wind-like profile in the case of eruptive origin. The opacity is adopted as 0.2 cm$^2$ g$^{-1}$. The free parameters and the priors are presented in Supplementary Table~\ref{tab: priors_fit}. As this supernova was detected with All-Sky Automated Survey for SuperNovae (ASAS-SN) \cite{2014ApJ...788...48S, 2018ApJ...853...57B} three days before discovery with Gaia, we set the epoch of first ASAS-SN detection ($t=-32$ d from maximum) as the upper limit for explosion date ($t_\mathrm{exp}$). Modeling each of post-peak bumps is essential to infer the total CSM mass and hence the progenitor properties. In order to fit all these features, we set conditions for initial time of ejecta-CSM collision,
\begin{eqnarray}
\left\{
\begin{array}{lll}
    t_{\mathrm{i},2}<t_\mathrm{b,2},\\
    t_\mathrm{b,2}<t_{\mathrm{i},3}<t_\mathrm{b,3},\\
    t_\mathrm{b,3}<t_{\mathrm{i},4}<t_\mathrm{b,4},
\end{array}
\right.
\label{eq: t_conditions}
\end{eqnarray}
where $t_{\mathrm{b},k=2,3,4}=123, 167, 219$ days represent the epochs of the observed post-peak bumps in the bolometric light curve.

The posterior distributions for these free parameters are obtained with the Markov chain Monte Carlo (MCMC) sampling package \texttt{emcee}\cite{2013PASP..125..306F}. In the maximum likelihood analysis, we introduce a weight function $\omega(t_i)$ and modify the log likelihood function as
\begin{equation}
\ln {\cal L} = - \frac{1}{2}\sum_j^{N} \left\{\frac{\left[O(t_j)-M(t_j)\right]^2}{\sigma(t_j)^2/\omega(t_j)^2 } + \ln\left[2\pi\sigma(t_j)^2/\omega(t_j)^2\right]\right\},
\label{eq: lnL}
\end{equation}
where $O(t)$, $\sigma(t)$, and $M(t)$ are the observed luminosity, the corresponding error, and the model-predicted luminosity at the phase $t_j$, respectively. Note that the t$\sim+250 - +350$ d observations are relatively poor sampling than the earlier phases, which could result in poor constraints on the nickel mass. Thus, in order to place better constraints on late-time power sources, we increase the weight of late-time data in likelihood function by assuming $\omega(t_j)=1$ for $t_j\leq +250$ d and $\omega(t_j)=2$ for $t_j>+250$ d. 

Following the above procedure, the resultant fitting parameters (mean value and $1\sigma$ error) are presented in the Extended Data Table~\ref{tab: Lum_fit}. Posterior distributions of a subset of key parameters are shown in Supplementary Fig.~\ref{fig: mcmc}.

Based on the fitting results, both the forward and reverse shock velocities are always faster than $1000$ km s$^{-1}$ during all four collisions. Assuming electron-ion energy equipartition, the electron temperature at the shock front is proportional to the square of shock velocity \cite{1982ApJ...258..790C}. Hence, the temperatures at both forward and reverse shock fronts are always above several $10^7$ K, which are high enough to ionize neutral oxygen with an ionization energy 13.618 eV (corresponding to an ionization temperature of $1.58\times10^5$ K).

\subsection{Spectroscopy}

We collected a total of 25 spectra of SN~2017egm (covering phases from $t \approx -22$\,days to $t \approx +380$\,days) with six instruments: the Beijing Faint Object Spectrograph and Camera (BFOSC) on the 2.16\,m telescope at Xinglong Observatory (XLT) of National Astronomical Observatories of the Chinese Academy of Sciences (NAOC), the Double Beam Spectrograph (DBSP) \cite{1982PASP...94..586O} on the 5\,m (200\,inch) Hale telescope at Palomar Observatory (P200), the Yunnan Faint Object Spectrograph and Camera (YFOSC) on the 2.4\,m Lijiang telescope (LJT) \cite{2015RAA....15..918F} at Yunnan Observatories, the Kast spectrograph \cite{1993Miller} on the 3\,m Shane telescope at Lick Observatory (Lick-3m), and the Low Resolution Imaging Spectrometer (LRIS) \cite{1995PASP..107..375O} and the Multi-Object Spectrometer for Infrared Exploration (MOSFIRE) \cite{2012SPIE.8446E..0JM} mounted on the Keck-I 10\,m telescope. Note that the Keck spectra were obtained under several proposals (principal investigators C. Fremling, M. Kasliwal, R. Foley, R. Lunnan and S. Kulkarni), where the spectral observations at $t=141$~d, $142$~d and $t=312$~d were collected from the publicly available database of Keck observatory archive. The XLT+BFOSC spectra were taken at low airmass; the Keck+LRIS spectra were observed at the parallactic angle with an atmospheric dispersion compensator; and the other spectral observations were performed with the slit oriented along the parallactic angle \cite{1982PASP...94..715F}. The above steps help alleviate the effects of atmospheric dispersion on our spectra. The journal of spectroscopic observations is given in Supplementary Table~4.

All of the spectra were reduced with standard \texttt{iraf} routines. Raw images were pre-processed following traditional steps, such as corrections for bias and flat field, and removal of cosmic rays. Then, the one-dimensional spectra were optimally extracted from the pre-reduced two dimensional frames. The wavelength calibration was performed using comparison-lamp spectra which were obtained with the same instrumental configuration as the SN on the same night. The flux calibration of the continuum was based on the spectrophotometric standard stars observed on the same night. The spectra were further corrected for atmospheric extinction of local observatories and (except in the case of DBSP) as well as possible for telluric absorption (H$_2$O and O$_2$). We also extracted a galaxy spectrum from a region far from the SN site in the LRIS data obtained on 13 January 2018 ($t=198.7$ d). Note that the slit avoids the nucleus of the host galaxy and the extraction region is presumably in a spiral arm. This spectrum is characterized by narrow lines of hydrogen, helium, oxygen and sulfur. A continuum flux excess at short wavelength is observed in both the host-galaxy spectrum and the supernova spectrum at $t=312.1$d, implying that the latter is strongly contaminated by host-galaxy light. The reduced spectra are shown in Extended Data Fig.~2. 

As shown in Extended Data Fig.~\ref{fig: NaI_HeI}, H$\alpha$ can be detected in SN from near-maximum to nebular phases, while He\,\textsc{i} emission and Na\,\textsc{i}\,D absorption features were absent in the pre-maximum spectra but can be identified in the late-time Keck+LRIS spectra. In the five LRIS spectra observed during $t=+141.5-374.1$~d, the profile of H$\alpha$ is roughly similar with full width at half maximum (FWHM) of $200-300$ km s$^{-1}$, while the velocity width of He\textsc{i} 5876\,\AA\, line varies between 200--500 km s$^{-1}$. It is worth noting that H$\alpha$, He\,\textsc{i} and Na\,\textsc{i}\,D features from SN spectra suffer from contamination of background light, as these feature are also detected in the host-galaxy spectrum with roughly similar width. The width variation of He\textsc{i} could result from contribution from variable helium-rich CSM, different observation conditions (seeing condition and slit orientation) or perhaps the choices of the region used to extract the spectrum in data-reduction process. 

The spectral evolution of SN~2017egm is typical of SLSNe-I. The early-time spectra of SN~2017egm are characterized by a hot blue continuum and a strong O\,\textsc{ii} absorption complex as well as weak features of C\,\textsc{ii} and O\,\textsc{i}. In the nebular phase, the supernova ejecta cool and become largely transparent. The corresponding spectra have prominent emission lines, such as O\,\textsc{i} $\lambda7774$, Mg\,\textsc{i}] $\lambda4571$, Ca\,\textsc{ii} H\&K, [Ca\,\textsc{ii}] $\lambda\lambda$7291, 7323, and the Ca\,\textsc{ii} near-infrared (NIR) triplet \cite{2019ApJ...871..102N}. Based on the studies of a large SLSN-I sample \cite{2017ApJ...835...13J, 2019ApJ...871..102N}, [Ca\,\textsc{ii}] $\lambda\lambda$7291, 7323 is likely blended with [O\,\textsc{ii}] $\lambda\lambda$7320, 7330; Mg\,\textsc{ii} $\lambda\lambda$7877, 7896 could contribute to the red wing of the O\,\textsc{i} $\lambda7774$ profile; and the emission features around 9200\,\AA\ could result from O\,\textsc{i} $\lambda9263$ and/or Mg\,\textsc{ii} $\lambda9218$. Unlike most SLSNe-I that show prominent nebular lines of [O\textsc{i}] $\lambda\lambda$6300, 6364, the 6300\,\AA\, feature of SN 20173gm is weak and appears to be asymmetric about the rest wavelengths of [O\textsc{i}] at $t = 120$--210\,d (Extended Data Fig.~\ref{fig: spec_OCa}), possibly owing to blends with other absorption or emission features. From $t=312$ days onwards, [O\textsc{i}] can be clearly identified in the spectra, but still much weaker than 7300\,\AA\, emission.

In the near-infrared spectrum obtained with Keck/MOSFIRE at $t = +193$\,d, a prominent emission line is detected at $\sim 1.08$\,$\mu$m, with a velocity width of $\sim 6300$\,km\,s$^{-1}$ (Extended Data Fig.~\ref{fig: comp_NIR}). This feature is coincident with He\,\textsc{i} $\lambda$10,830, and its velocity width is comparable to that of H$\alpha$ emission lines detected in the late-time spectra of iPTF13ehe and iPTF15esb ($4000-6000$ km s$^{-1}$) \cite{2015ApJ...814..108Y, 2017ApJ...848....6Y}, suggesting a possibility that the broad emission at 1.08 $\mu$m could arise from the fast-moving helium-containing CSM that has been accelerated by previous ejecta-CSM interaction. Such a feature is also seen in the superluminous SN~2017gci and the normal Type Ic supernova SN~2013ge, though with a broader line profile. This feature in SN~2017gci is likely attributed to He, Mg or S lines \cite{2021MNRAS.502.2120F}, while for Type Ic supernovae, it is usually interpreted as C\,\textsc{i} and Mg\,\textsc{ii} \cite{2022ApJ...925..175S}. Note that SN~2013ge contains little helium as its early-time spectra appear to show weak lines of He\,\textsc{i} at optical and near-infrared wavelengths \cite{2016ApJ...821...57D}. Thus, this NIR feature of SN~2017egm is likely to be a blend of multiple elements including He\,\textsc{i}. The presence of some helium in SN 2017egm is also favored by a recent study with additional NIR data\cite{2023arXiv230303424Z}.

\subsection{Host-Galaxy Extinction}

At nebular phases when the supernova was quite faint, the Keck+LRIS spectra show a clear Na\,\textsc{i}\,D absorption feature (Extended Data Fig.~\ref{fig: NaI_HeI}), corresponding to $E(B-V) \approx 0.2$\,mag based on the statistical correlation between host extinction and equivalent width of Na\,\textsc{i}\,D \cite{2012MNRAS.426.1465P}. Moreover, based on the Balmer decrement method, the intensity ratios of H\,\textsc{i} in the $t = 374.1$\,d spectrum also suggest a high host extinction of $E(B-V) \approx 0.25$\,mag. However, the absence of narrow Na\,\textsc{i}\,D absorption in our early-time spectra indicates low host extinction at supernova location \cite{2018ApJ...853...57B}, which is also supported by the low H\,\textsc{i} column density along the line of sight \cite{2018ApJ...858...91Y}. Thus, SN~2017egm could be located in the outskirt of its host galaxy, NGC~3191. Here the contribution of the internal extinction is neglected.

\subsection{Environmental Metallicity}

Extended Data Fig.~\ref{fig: galaxy} shows the identification of the host-galaxy emission lines in the Keck spectrum obtained on July 13 2018 ($t = +374.1$\,d) at the position of SN~2017egm. The spectral ranges over $7200 < \lambda < 7500$\,\AA\ and $\lambda > 8390$\,\AA\ are excluded from further analysis of the supernova environment, owing to contamination of nebular oxygen and calcium emission from the SN ejecta. The spectral synthesis code \texttt{STARLIGHT} \cite{2005MNRAS.358..363C} was then used to extract the stellar component of the host galaxy from the observed spectrum. The prominent host emission lines include H\,\textsc{i}, [N\,\textsc{ii}], [O\,\textsc{ii}], [O\,\textsc{iii}], and [S\,\textsc{ii}]. For H$\beta$, [O\,\textsc{ii}], and [O\,\textsc{iii}], we applied a Gaussian fit to the continuum-subtracted line profile. A double Gaussian function is used to fit the [S\,\textsc{ii}] doublets, while a triple Gaussian fit is performed to the profile of H$\alpha$ and [N\,\textsc{ii}] $\lambda\lambda$6548, 6584.

The flux ratios of strong emission lines can be used to determine the metallicity of the galaxy according to empirical relations established between oxygen abundance and the ratios. As seen in Extended Data Fig.~\ref{fig: metallicity}, the inferred metallicities from various diagnostics are 12 + log(O/H) = 8.53--9.12, corresponding to 0.7--2.7\,Z$_\odot$ (assuming a solar oxygen abundance of 8.69 \cite{2009ARA&A..47..481A}), in agreement with the previous literature \cite{2017ApJ...849L...4C, 2017ApJ...845L...8N, 2018ApJ...853...57B, 2018A&A...610A..11I, 2018ApJ...858...91Y}. However, the results of most diagnostics support a high metallicity (1.2--2.7\,Z$_\odot$), suggesting that SN~2017egm likely originated in a metal-rich environment, contrary to the trend that most SLSNe-I reside in low-metallicity hosts \cite{2014ApJ...787..138L, 2015MNRAS.449..917L, 2016ApJ...830...13P, 2018MNRAS.473.1258S}.

\subsection{Mass-Loss Rate}
Based on the properties of CSM inferred from the MCSIRD-powered model fit, the eruption responsible for the fourth CSM shell started at $T_\mathrm{w}=R_{\mathrm{CSM,out},4}/v_\mathrm{w}$=2$/(v_\mathrm{w}/1000$ km s$^{-1}$) years before core collapse, where $R_{\mathrm{CSM,out},4}=3[M_{\mathrm{CSM},k}(3-s)/(4\pi \rho_{\mathrm{CSM},k})+ R_{\mathrm{CSM,in},4}^3]^ {1/3}$ is the outer radius of the fourth shell and $v_\mathrm{w}$ is the shell velocity. Within such a short duration, the progenitor of SN~2017egm ejected a total mass of 6.8--7.7~M$_\odot$. The average mass-loss rate during this period is about $\dot{M}\sim 4(v_\mathrm{w}/1000$ km s$^{-1}$)~M$_\odot$ yr$^{-1}$. See main text for more discussions.

\subsection{Comparison with normal type Ic supernovae}

A star characterized by such an intense mass-loss history is reminiscent of a Wolf-Rayet (WR) star that evolved from an eruptive LBV. It is interesting to consider whether the progenitor of SN~2017egm could be a higher-mass extension of the hydrogen/helium-free progenitors of lower-luminosity type Ibc supernovae. A high environmental metallicity (1.2--2.7\,Z$_\odot$), favored by most of our measurements based on various host-line diagnostics, is typical of the ordinary type Ic supernovae hosts but significantly higher than the abundance distribution inferred for most SLSNe-I \cite{2014ApJ...787..138L, 2015MNRAS.449..917L, 2016ApJ...830...13P, 2021ApJS..255...29S}. We also note that the inferred $^{56}$Ni mass (0.07\,M$_\odot$) from light-curve modelling for SN~2017egm is comparable to those of type Ic supernovae ($\sim0.05$--0.8\,M$_\odot$) \cite{2016MNRAS.458.2973P}. While the overall population properties (metallicity and $^{56}$Ni mass) are similar to those of SNe~Ib/c, the large amount of CSM is not expected from even the massive end of type Ibc supernovae progenitors, which usually have much lower mass loss. 

\subsection{Comparison with iPTF14hls}

In analogy with SN~2017egm, the long-lived Type II-P supernova SN iPTF14hls shows complex luminosity undulations \cite{2017Natur.551..210A, 2019A&A...621A..30S} (that is, spiked bumps and sharp turns of different decline slopes), although the latter has much broader profile of bumps (see inset of right panel of Extended Data Fig.~\ref{fig: LTR_sne}). This supernova experienced extremely slow spectral evolution over 1000 days, about 10 times slower than its peers  \cite{2017Natur.551..210A, 2019A&A...621A..30S}. The spectra of SN~2017egm are characterized by absorption features before the rapid decline of luminosity at $t \approx 120$\,d, while emission features dominate and remain largely unchanged during the undulations of $t=+150$ to +210\,d (see Extended Data Figs.~\ref{fig: spec} and \ref{fig: spec_OCa}). Thus, both the light curves and spectral evolution of the two supernovae require multiple/long-term strong energy input. In the interaction-powered scenario, such various bumps possibly indicate a non-periodic frequent intense mass-loss history for these two SNe. The PPI-related mechanisms (e.g., shell-shell interaction or ejecta-shell interaction) have been also discussed in the case of iPTF14hls \cite{2017Natur.551..210A, 2019A&A...621A..30S, 2022ApJ...933..102W}, but they have difficulties in interpreting the continuous presence of hydrogen in iPTF14hls \cite{2017Natur.551..210A}. For the hydrogen-poor SN~2017egm, interaction between the supernova ejecta and PPI-driven shells seems to be a very likely explanation for its bumpy light curves.

\section*{Data availability}
All data of SN~2017egm reported in this paper are publicly available via the Figshare repository (https://doi.org/10.6084/m9.figshare.22009868) \cite{SN2017egmdataset} and the Weizmann Interactive SN Repository (https://wiserep.weizmann.ac.il) \cite{2012PASP..124..668Y}. The photometry is also presented in Supplementary material.

 \section*{Acknowledgements}
 We acknowledge the staff of the 50BiN, AZT, LJT, {\it Swift}, XLT, Lick, Palomar, and Keck Observatories for assistance with the observations. We thank R. J. Foley and S. Kulkarni for their Keck+LRIS observations, and S. Adams for taking the Keck+MOSFIRE spectrum. We also thank J. Sollerman for sharing the luminosity data of iPTF14hls, S. Saito for providing Subaru spectrum of SN 2017egm, L. Dessart and A. Jerkstrand for useful discussions, and M. Nicholl and Q. Fang for providing their results of spectral analysis for normal core-collapse and/or superluminous supernovae. The operation of XLT is supported by the Open Project Program of the Key Laboratory of Optical Astronomy, National Astronomical Observatories, Chinese Academy of Sciences. Funding for the LJT has been provided by the Chinese Academy of Sciences and the People's Government of Yunnan Province. The LJT is jointly operated and administrated by Yunnan Observatories and Center for Astronomical Mega-Science, CAS. This research has made use of the Keck Observatory Archive (KOA), which is operated by the W. M. Keck Observatory and the NASA Exoplanet Science Institute (NExScI), under contract with the National Aeronautics and Space Administration. The W. M. Keck Observatory is operated as a scientific partnership among the California Institute of Technology, the University of California, and NASA; the observatory was made possible by the generous financial support of the W. M. Keck Foundation. The Kast red CCD detector upgrade, led by B.  Holden, was made possible by the Heising-Simons Foundation, William and Marina Kast, and the University of California Observatories. Research at Lick Observatory is partially funded by a generous gift from Google.

The work of X.W. is supported by the National Natural Science Foundation of China (NSFC grants 12288102 and 12033003), the Scholar Program of Beijing Academy of Science and Technology (DZ:BS202002), and the Tencent Xplorer Prize. A.G.-Y.'s research is supported by the EU via ERC grant 725161, the ISF GW excellence center, an IMOS space infrastructure grant and BSF/Transformative and GIF grants, as well as the Andr\'{e} Deloro Institute for Advanced Research in Space and Optics, the Schwartz/Reisman Collaborative Science Program and the Norman E. Alexander Family Foundation ULTRASAT Data Center Fund, Minerva and Yeda-Sela; A.G.-Y. is the incumbent of the The Arlyn Imberman Professorial Chair. A.V.F.'s group at UC Berkeley received financial support from the Miller Institute for Basic Research in Science (where A.V.F. was a Miller Senior Fellow), the Christopher R. Redlich Fund, and many individual donors. N.B. is funded/co-funded by the European Union (ERC, CET-3PO, 101042610). Views and opinions expressed are however those of the author(s) only and do not necessarily reflect those of the European Union or the European Research Council. Neither the European Union nor the granting authority can be held responsible for them. L.X. acknowledges support from National Natural Science Foundation of China (grant no. 12103050), Advanced Talents Incubation Program of the Hebei University, and Midwest Universities Comprehensive Strength Promotion project. L.W. acknowledges support from the National Program on Key Research and Development Project of China (grant 2021YFA0718500).

 \section*{Author contributions}
W.L. and X.W. wrote the paper. W.L. led the analysis. X.W. led the overall project. L.Y. and A.G.-Y. assisted with the paper. J.M. reduced the 50BiN, AZT and TNT data. T.G.B. reduced the Keck LRIS data. L.Y. obtained and reduced P200 telescope data. A.V.F., T.G.B. and W.Z. obtained and reduced the Lick telescope data. A.V.F. also helped with the manuscript. D.X. and H.L. reduced the XLT data. R.L. advised on the interpretation of narrow spectral lines. P.B. reduced the Swift data. M.K. and L.Y. provided the MOSFIRE data. R.L., C.F. and N.B. provided their LRIS observations. D.M. and S.A.E. obtained the AZT imaging. K.Z. and Jicheng Z. obtained the XLT observations. S.Y. and Jujia Z. reduced the LJT data. Z.C., L.D., K.W. and L.X. helped with the draft. L.W. helped check the model code.

\section*{Competing interests}
The authors declare no competing interests.

\section*{Additional information}
{\bf{Correspondence and requests for materials}} should be addressed to W.L. and X.W..

\section*{Figure Legends/Captions}
\clearpage

\clearpage

\renewcommand{\figurename}{Extended Data Figure} \renewcommand{\tablename}{Extended Data Table} 
\setcounter{figure}{0}

\begin{figure}
\center
\includegraphics[angle=0,width=0.5\textwidth]{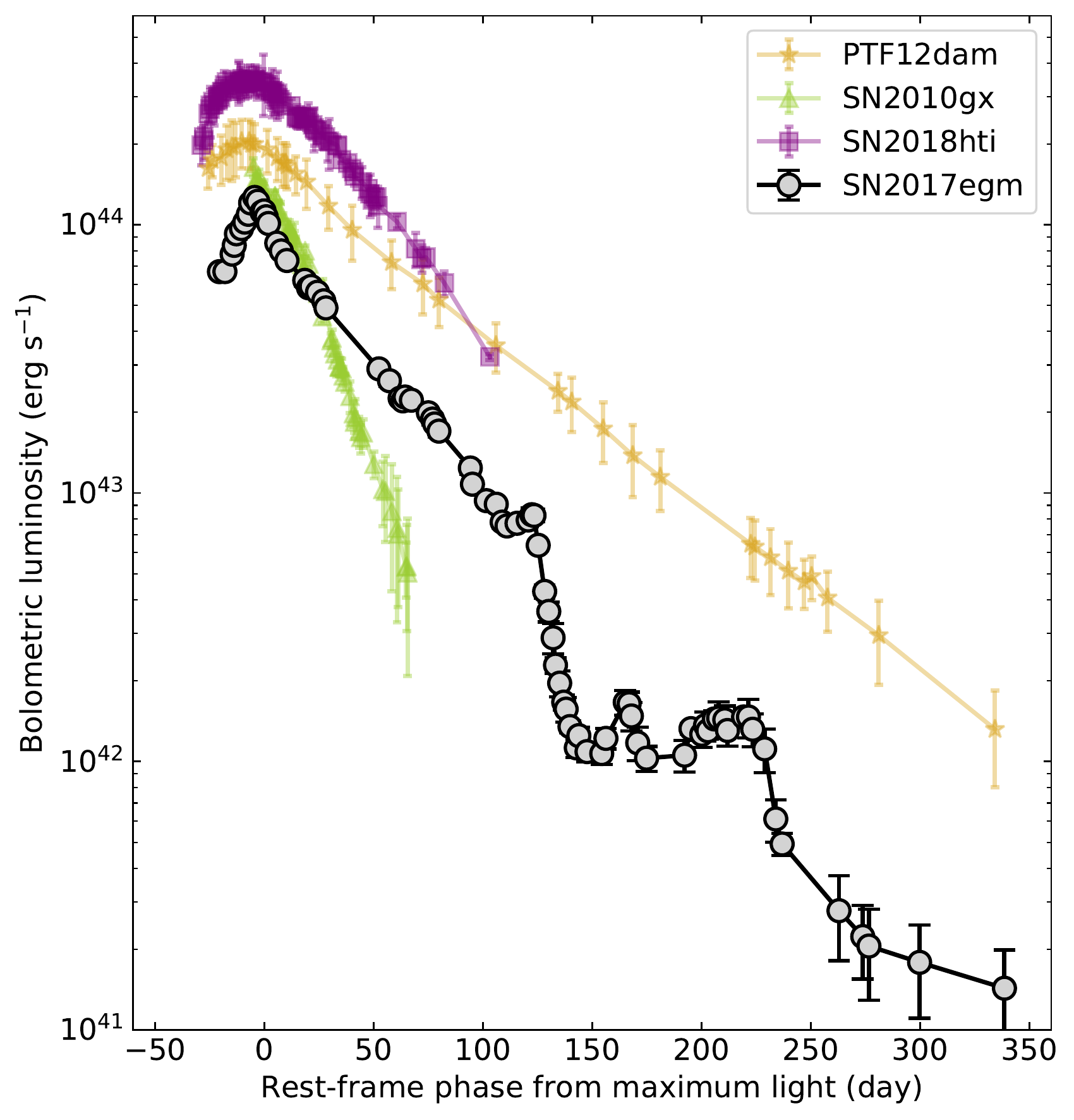}\includegraphics[angle=0,width=0.5\textwidth]{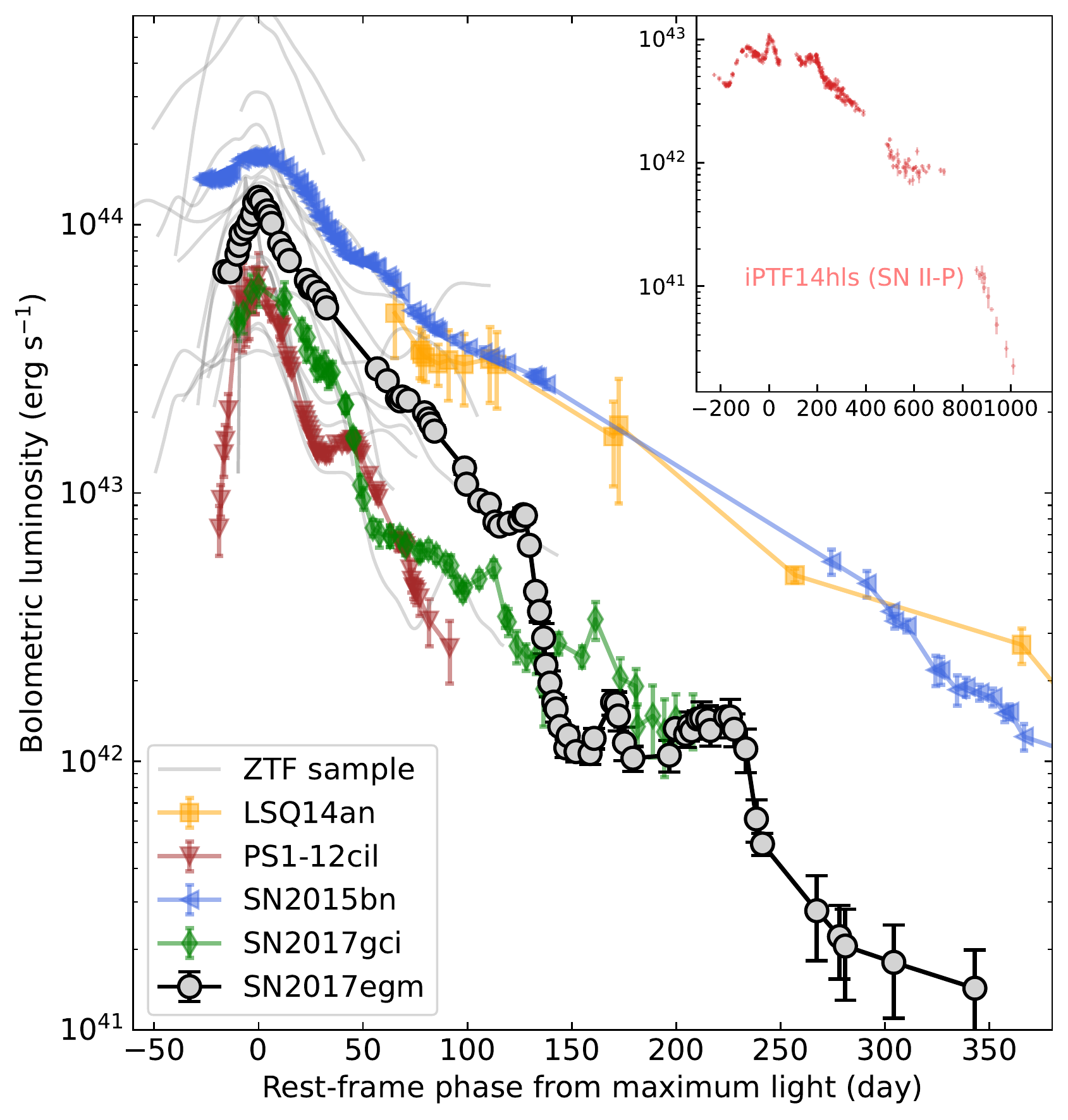}
\caption{{\bf{The bolometric light curve of SN~2017egm, as compared to some representative SLSNe-I.}} Comparison SLSNe-I include objects with monotonic decay after maximum light ({\it left panel:} PTF12dam, yellow stars; SN~2010gx, yellow-green triangles; SN~2018hti, purple squares) and those with post-peak bumps ({\it right panel:} LSQ14an, orange squares; PS1-12cil, brown triangles; SN~2015bn, blue triangles; SN~2017gci, green diamonds; ZTF sample, gray lines). The bolometric luminosities of PTF12dam, SN~2017gci, SN~2018hti and ZTF sample are retrieved from the literatures\cite{2017ApJ...835...58V, 2020MNRAS.497..318L, 2021MNRAS.502.2120F, 2023ApJ...943...42C}, while those of other comparison SLSNe-I are derived by fitting the UV-absorbed blackbody curve \cite{2017ApJ...850...55N} to its multiband photometry \cite{2014Ap&SS.354...89B, 2016ApJ...826...39N, 2016ApJ...828L..18N, 2017MNRAS.468.4642I, 2018ApJ...860..100D, 2018ApJ...852...81L}. In the inset of the {\it right panel}, the luminosity evolution of a peculiar SN II iPTF14hls (red pluses) is shown for comparison \cite{2017Natur.551..210A, 2019A&A...621A..30S}. The error bars show 1$\sigma$ uncertainties. }
\label{fig: LTR_sne}
\end{figure}

\begin{figure}
\center
\includegraphics[angle=0,width=0.6\textwidth]{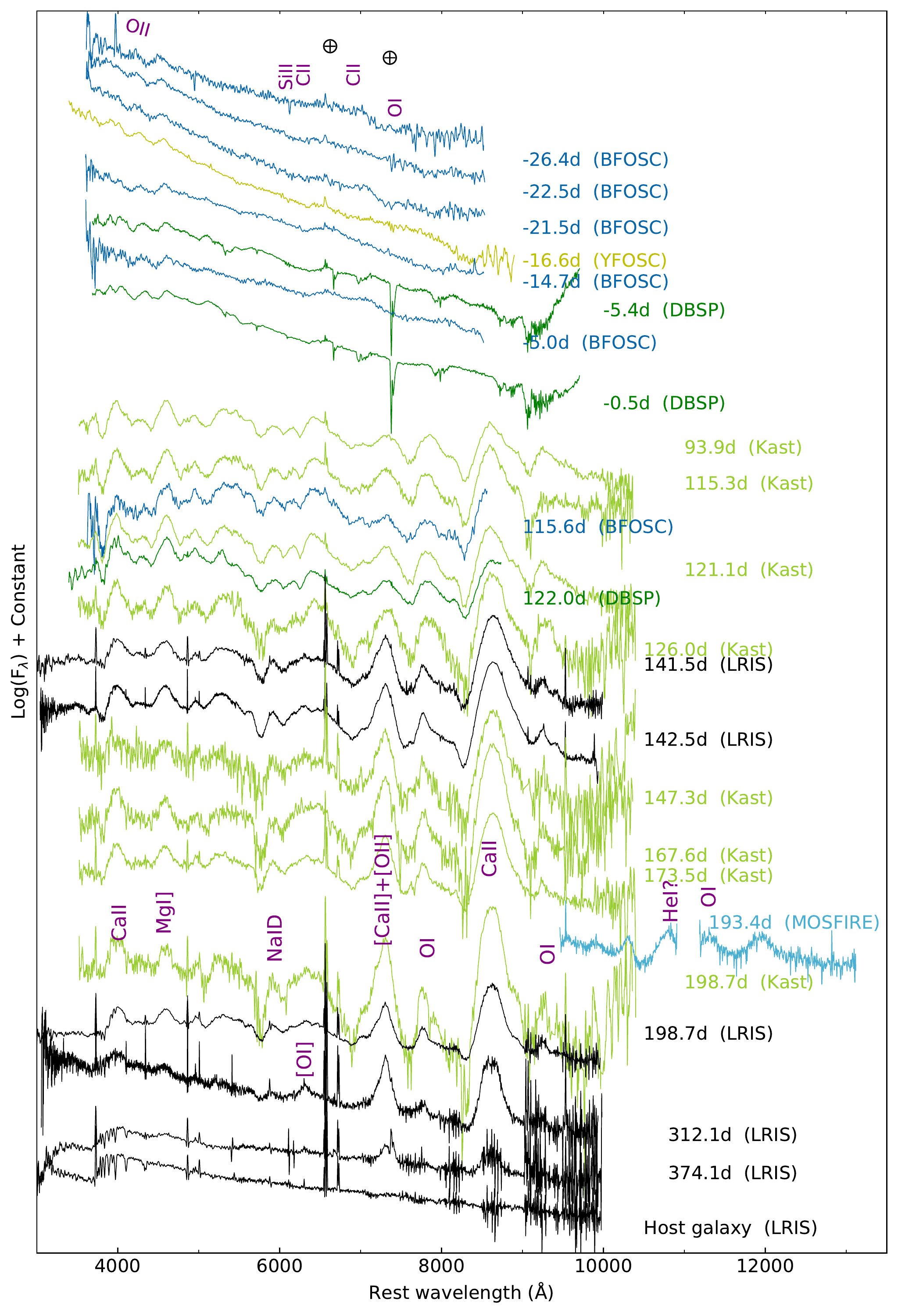}
\caption{{\bf{Spectra of SN~2017egm and its host galaxy.}} The observations were obtained with Lick+Kast (yellow green), LJT+YFOSC (yellow), Keck+LRIS (black), Keck+MOSFIRE (light blue), P200+DBSP (green), and XLT+BFOSC (blue). All of the spectra were smoothed with a Savitzky-Golay filter. Telluric absorption that was not removed in some of the spectra is marked with an Earth symbol.}
\label{fig: spec}
\end{figure}

\begin{figure}
\center
\includegraphics[angle=0,width=0.8\textwidth]{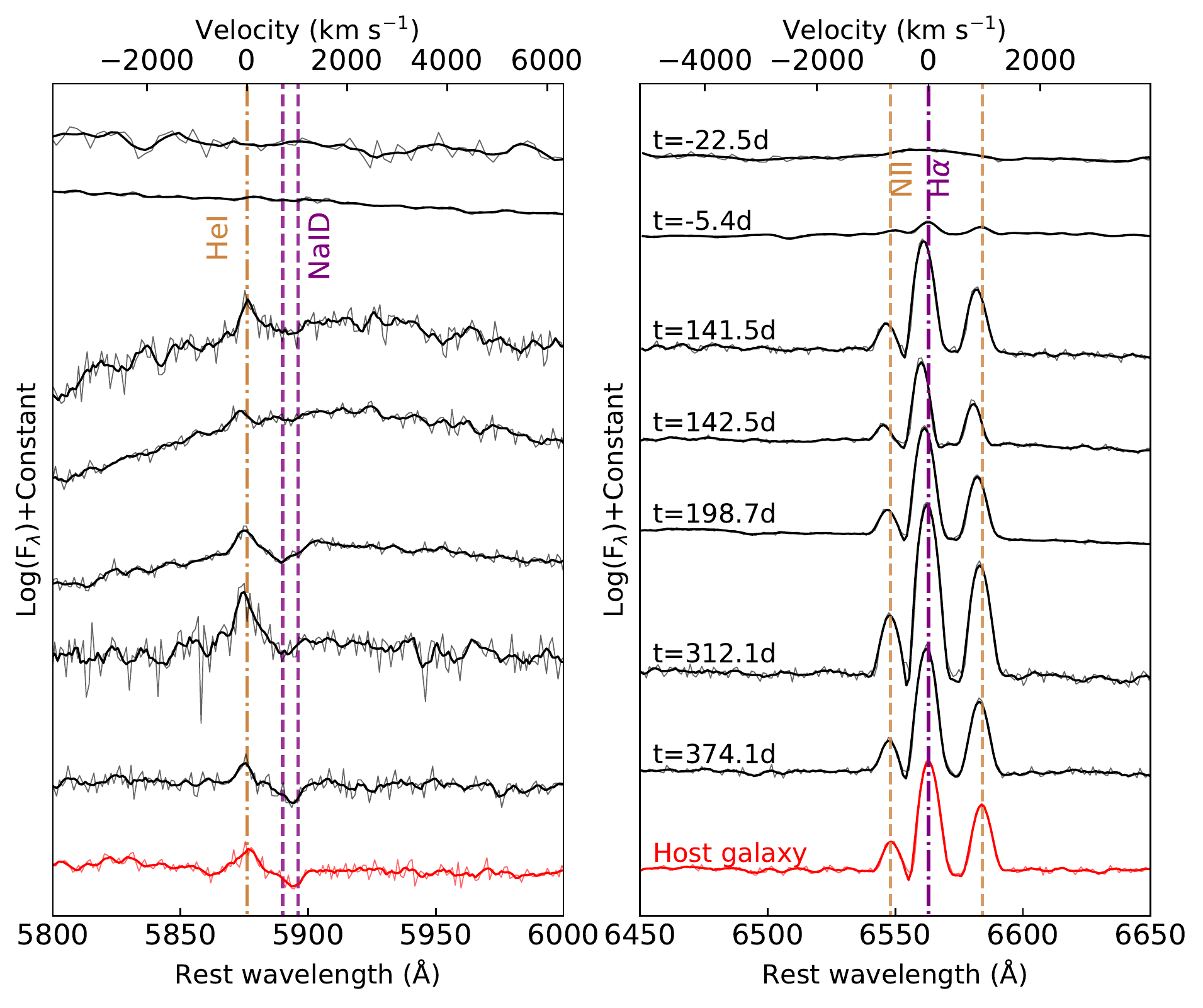}
\caption{{\bf{Narrow spectral features observed in the spectra of SN~2017egm and its host galaxy.}} The spectra of SN and galaxy are displayed in black and red, respectively. {\it Left panel:} He\,\textsc{i} emission (yellow dot-dashed) and Na\,\textsc{i}\,D absorption (purple dashed). {\it Right panel:} [N\,\textsc{ii}] (yellow dashed) and H$\alpha$ (purple dot-dashed).}
\label{fig: NaI_HeI}
\end{figure}

\begin{figure}
\center
\includegraphics[angle=0,width=\textwidth]{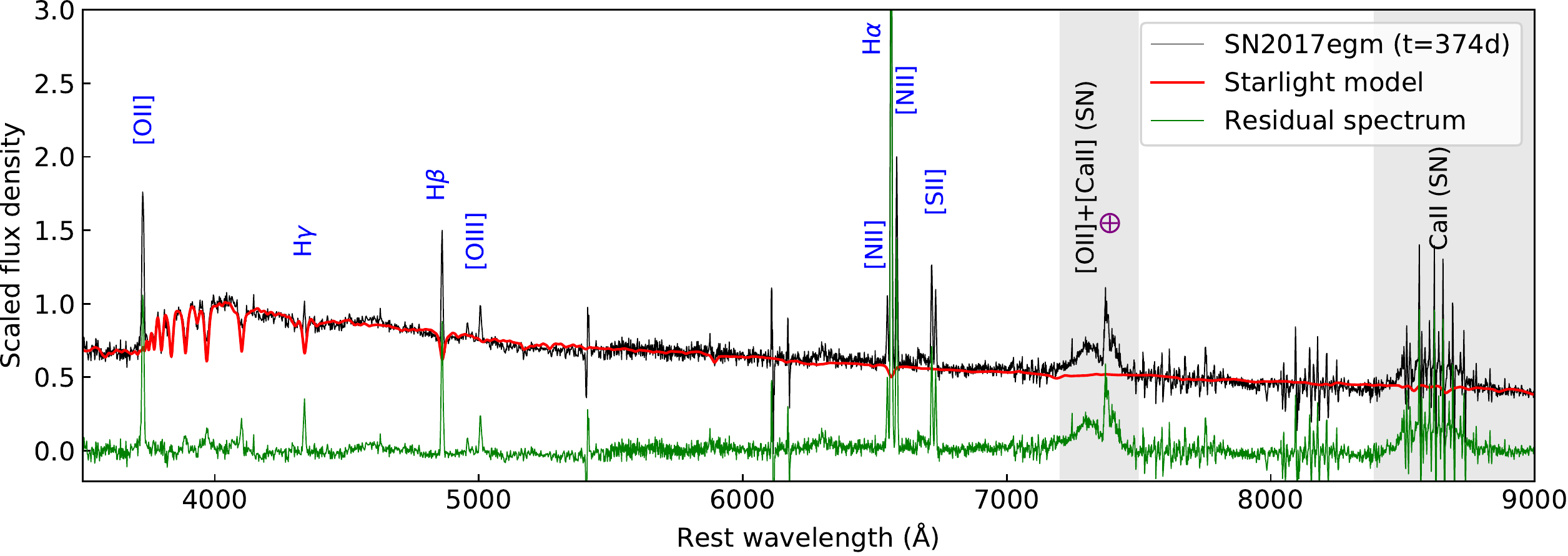}
\caption{{\bf{Identification of emission lines from the host galaxy of SN~2017egm}}. The black line shows the Keck+LRIS spectrum obtained at the SN location on 2018 July 13, and the green line displays the residual spectrum after subtraction of the host-galaxy stellar component (red lines) obtained with a starlight-fitting tool \cite{2005MNRAS.358..363C}. The oxygen and calcium emission features from the SN are shown in gray shade. Incompletely removed telluric contamination is marked with an Earth symbol.} 
\label{fig: galaxy}
\end{figure}

\begin{figure}
\center
\includegraphics[angle=0,width=0.5\textwidth]{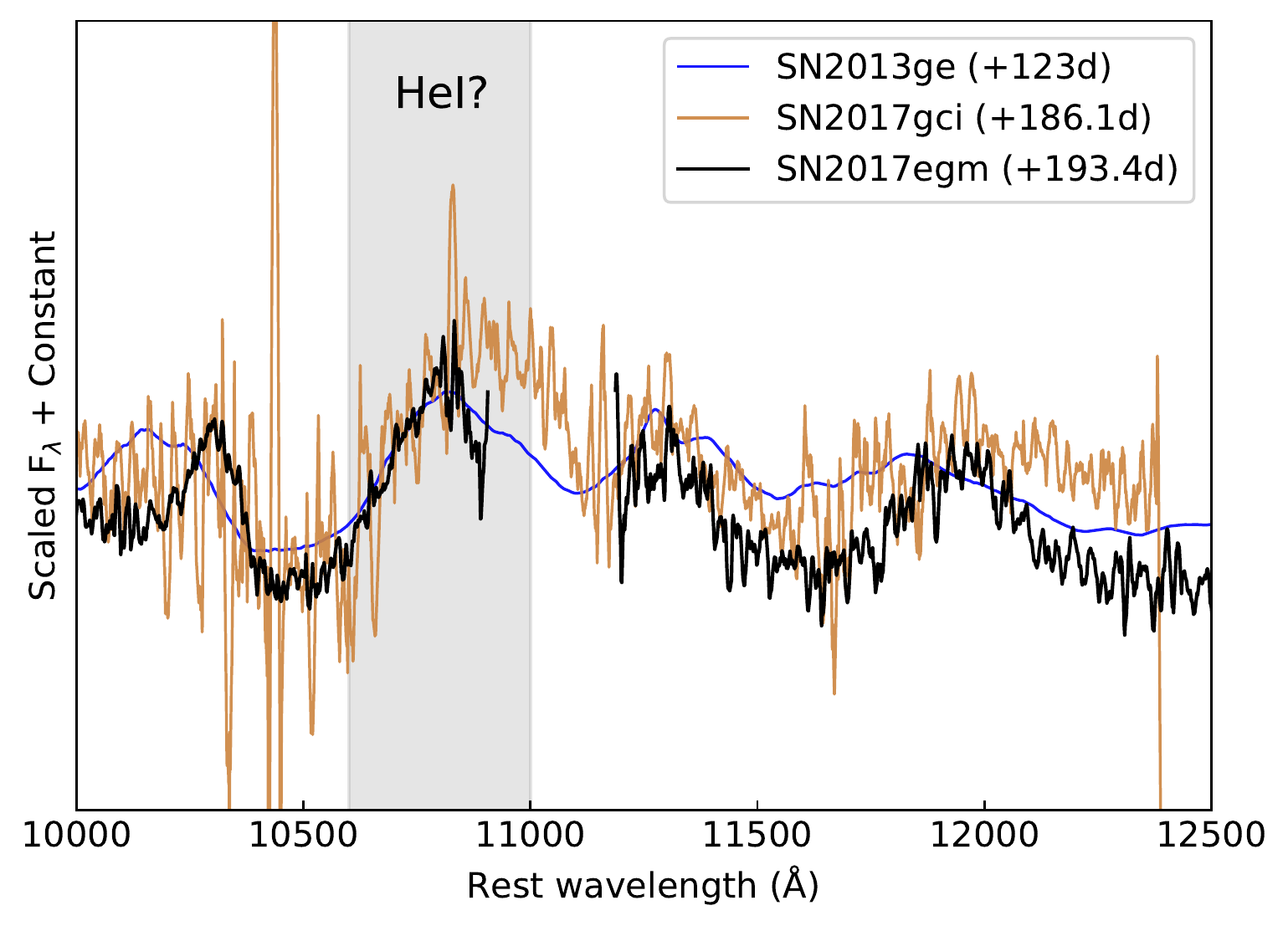}
\caption{{\bf{Near-infrared (NIR) spectrum of SN~2017egm}}. The NIR spectra of superluminous SN~2017gci (yellow) \cite{2021MNRAS.502.2120F} and Type Ic SN~2013ge (blue) \cite{2016ApJ...821...57D} are overplotted for comparison. The prominent feature coincident with He\,\textsc{i} $\lambda$10,830 is shaded in gray. All spectra were smoothed with a Savitzky-Golay filter. See text for discussion.}
\label{fig: comp_NIR}
\end{figure}

\begin{figure}
\center
\includegraphics[angle=0,width=0.8\textwidth]{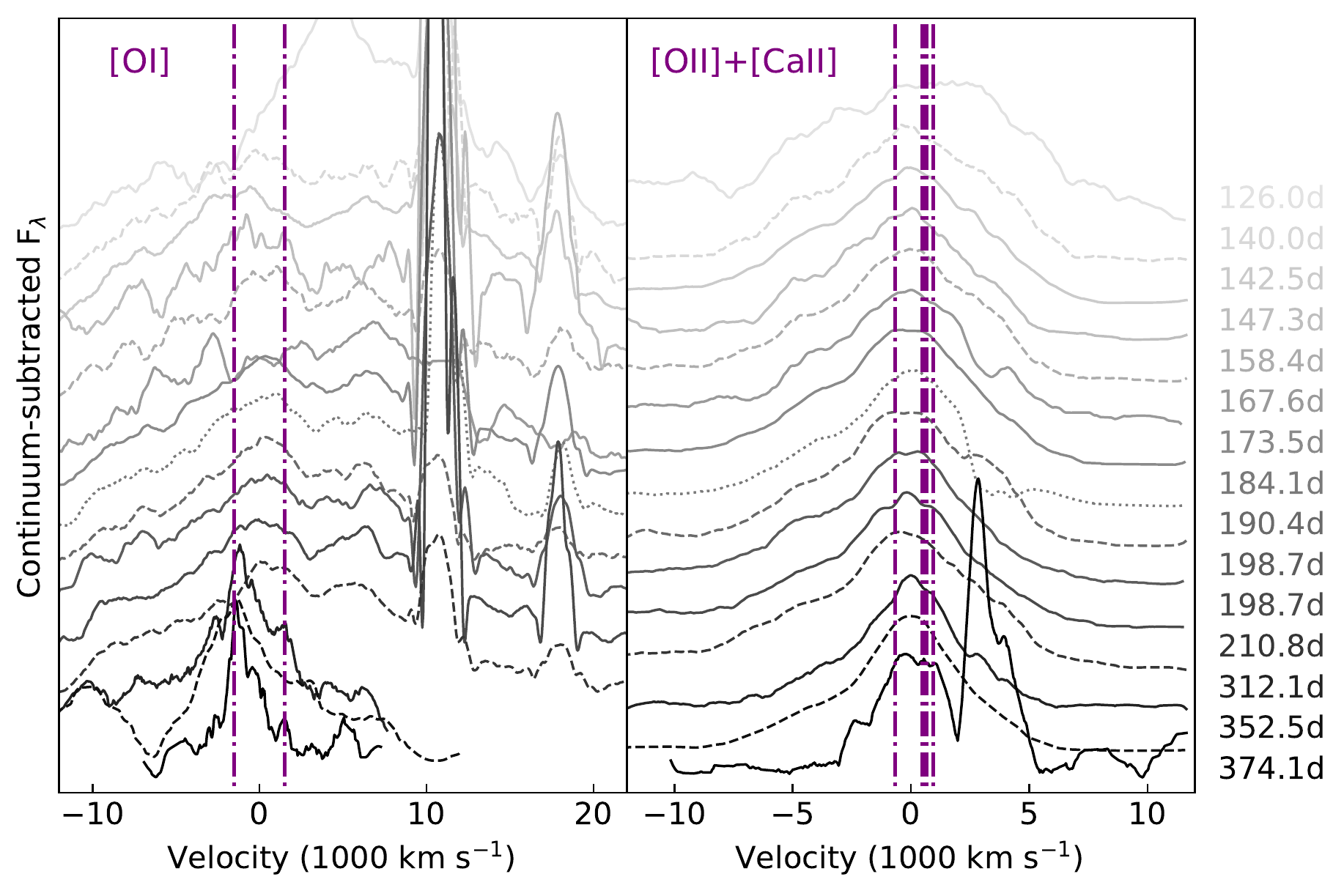}
\includegraphics[angle=0,width=0.8\textwidth]{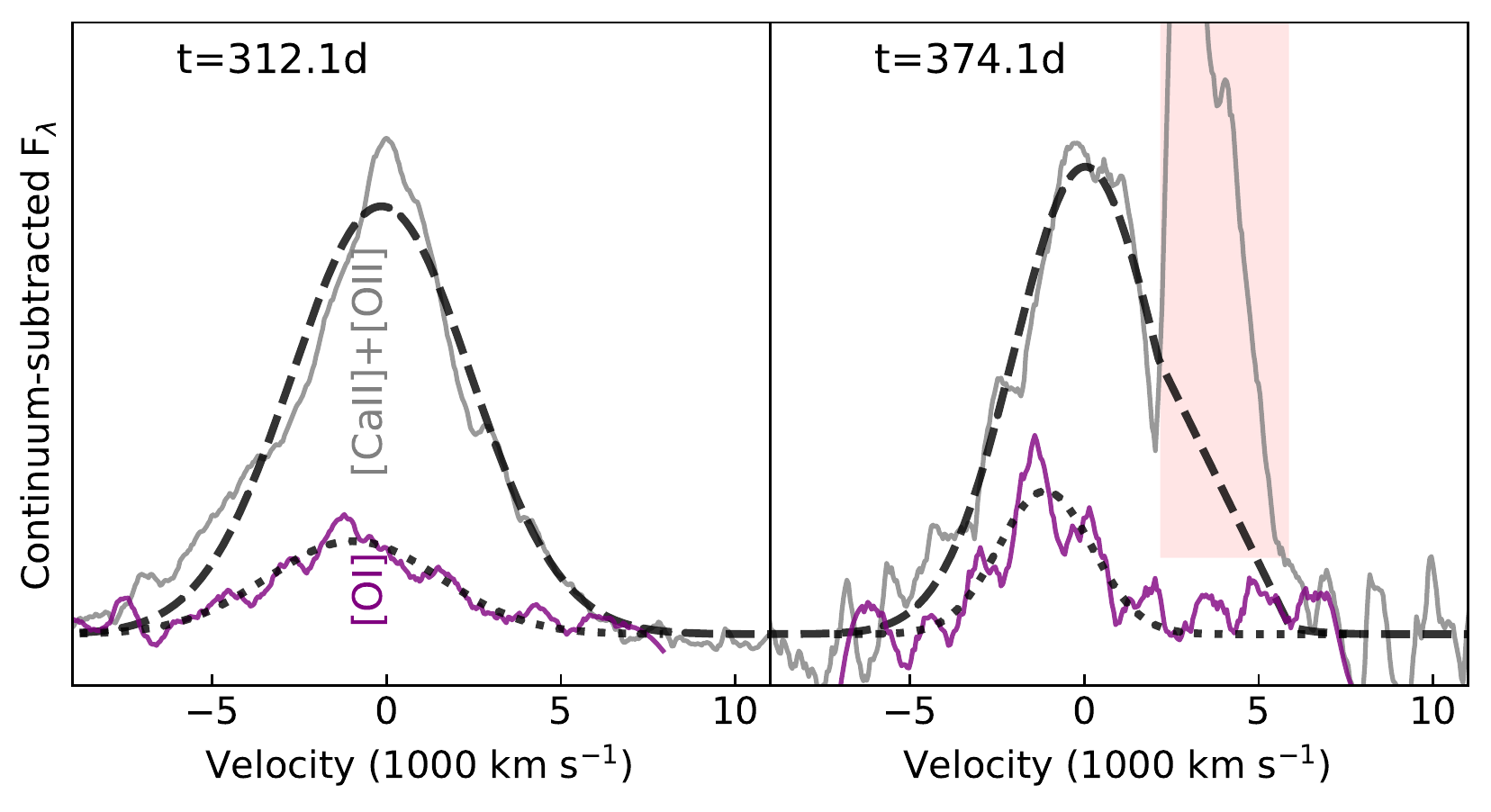}
\caption{{\bf{Close-up view of the [O\textsc{i}] $\lambda\lambda$6300, 6364 and [O\,\textsc{ii}] $\lambda\lambda$7319, 7330 + [Ca\textsc{ii}] $\lambda\lambda$7291, 7323 features in late-time spectra of SN~2017egm.}} {\it Upper panel:} observations of [O\textsc{i}] {\it (left)} and [O\,\textsc{ii}] + [Ca\textsc{ii}] {\it (right)} features during the interval $t = 126$--374\,d are shown with dashed \cite{2019ApJ...871..102N}, dotted \cite{2020ApJ...894..154S} and solid (this work) lines, respectively. The rest-frame wavelengths of these spectral lines are indicated by vertical dot-dashed lines. The 6300\,\AA\, feature at $t = 120$--210\,d appears to be asymmetric about the rest wavelengths of [O\textsc{i}], possibly owing to blends with other absorption or emission features. {\it Lower panel:} the photometry-calibrated profiles of [O\textsc{i}] (purple) and [O\,\textsc{ii}] + [Ca\textsc{ii}] (gray) at $t=312$\,d and $t=374$\,d can be well fitted with Gaussian functions, as indicated by the dotted and dashed profiles, respectively. The spectral coverage contaminated by the incompletely removed telluric line is shaded in red and excluded from the Gaussian fit. All spectra were smoothed with the Savitzky-Golay filter.}
\label{fig: spec_OCa}
\end{figure}

\begin{figure}
\center
\includegraphics[angle=0,width=0.5\textwidth]{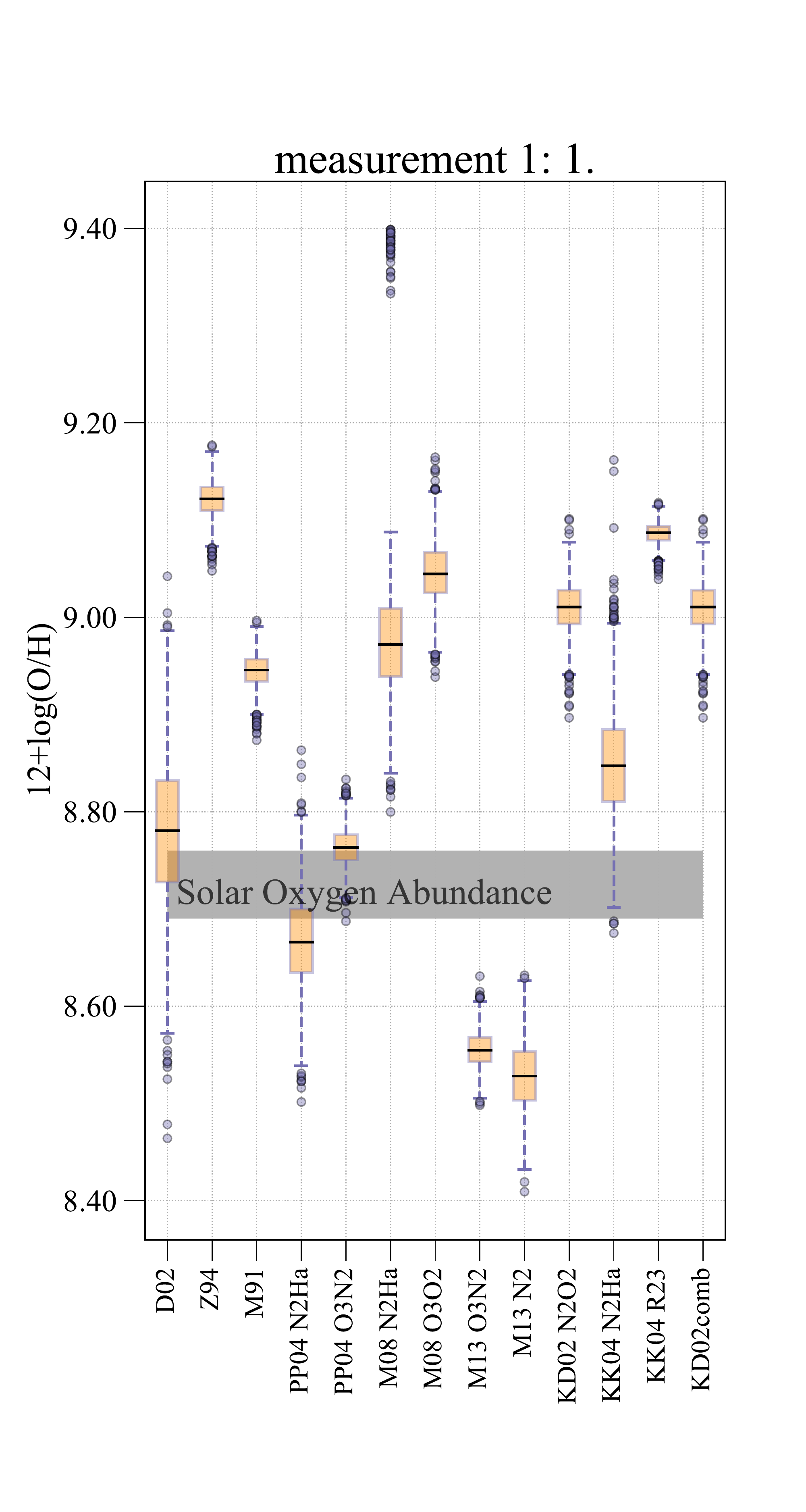}
\caption{{\bf{Environmental metallicity of SN~2017egm estimated from different line-ratio diagnostics via the \texttt{PYMCZ} code}} \cite{2016A&C....16...54B}. For each diagnostic, we perform 2000 Monte Carlo sampling to obtain the median (black solid line) and interquartile range (IQR; orange box) of the metallicity distribution; the distribution spanning the range of $1.5 \times$ IQR is represented by blue dashed lines (whisker), beyond which the sampling points are considered to be outliers; the minima and maxima values are shown by horizontal blue lines. The gray-shaded region shows the range of solar oxygen abundances reported in the literature.}
\label{fig: metallicity}
\end{figure}

\clearpage

\begin{table}
\center
\caption{{\bf{Fitting Parameters of MCSIRD Model for SN~2017egm}}. Here $t_\mathrm{exp}$ is the explosion date from maximum light; $x_0$ is the dimensionless transitional radius between inner and outer part of ejecta; $E_{\mathrm{SN}}$ and $M_{\mathrm{ej}}$ are the kinetic energy and mass of the SN ejecta; $M_\mathrm{Ni}$ is the newly-synthesised $^{56}$Ni mass; $t_{i,k}$ is the initial time of $k$-th interaction; $M_\mathrm{CSM,k}$ is the mass of $k$-th CSM shell, $\rho_\mathrm{in,k}$ is the CSM density at the inner radius ($R_\mathrm{in,k}$), and $\varepsilon_{\mathrm{fs},k}$ and $\varepsilon_{\mathrm{rs},k}$ are the efficiencies of converting forward and reverse shock energy to radiation. Here the subscript $k$ ($=1,2,3,4$) denotes the sequential number of CSM shell.}
\begin{tabular}{ccccc}
\hline\hline
\multicolumn{5}{c}{\bf{SN ejecta}}   \\\hline
$t_\mathrm{exp}$ &$x_0$& $E_\mathrm{SN}$ &  $M_\mathrm{ej}$ & $M_\mathrm{Ni}$   \\
(day)&  &($10^{51}$\,erg) &  ($M_\odot$) & ($M_\odot$) \\
\cline{1-5}\\
$-42.7_{-0.6}^{+0.4}$ & $0.74_{-0.03}^{+0.03}$& $0.876_{-0.018}^{+0.025}$ &$2.55_{-0.28}^{+0.38}$   & $0.07_{-0.03}^{+0.04}$ \\
\hline\hline
 \multicolumn{5}{c}{\bf{1st CSM shell}} \\
\cline{1-5}\\
$M_\mathrm{CSM,1}$ & $R_\mathrm{in,1}$ &$\rho_\mathrm{in,1}$& $\varepsilon_{\mathrm{fs},1}$ &$\varepsilon_{\mathrm{rs},1}$  \\
 ($M_\odot$)& ($10^{15}$\,cm) & ($10^{-14}$\,g\,cm$^{-3}$)\\
\cline{1-5}\\
$4.09_{-0.25}^{+0.23}$& $0.087_{-0.033}^{+0.037}$  & $26.1_{-1.9}^{+2}$ &$0.559_{-0.063}^{+0.095}$ & $0.083_{-0.059}^{+0.119}$ \\
\hline
\hline
 \multicolumn{5}{c}{\bf{2nd CSM shell}} \\
\cline{1-5}\\
$M_\mathrm{CSM,2}$  & $t_{i,2}$   &$\rho_\mathrm{in,2}$     &$\varepsilon_{\mathrm{fs},2}$  &$\varepsilon_{\mathrm{rs},2}$  \\
 ($M_\odot$)& (day) & ($10^{-14}$\,g\,cm$^{-3}$)\\
\cline{1-5}\\
$1.23_{-0.04}^{+0.05}$& $19.1_{-0.4}^{+0.6}$ & $13.7_{-1.1}^{+1.2}$ &$0.194_{-0.017}^{+0.019}$ & $0.955_{-0.057}^{+0.032}$\\
\hline
\hline
 \multicolumn{5}{c}{\bf{3rd CSM shell}} \\
\cline{1-5}\\
$M_\mathrm{CSM,3}$ & $t_{i,3}$   &$\rho_\mathrm{in,3}$    &$\varepsilon_{\mathrm{fs},3}$  &$\varepsilon_{\mathrm{rs},3}$   \\
 ($M_\odot$)& (day) & ($10^{-14}$\,g\,cm$^{-3}$)\\
\cline{1-5}\\
$0.85_{-0.07}^{+0.07}$  & $154_{-0.2}^{+0.2}$& $50.5_{-7.6}^{+7.7}$& $0.01$ & $0.027_{-0.006}^{+0.006}$ \\
\hline
\hline
 \multicolumn{5}{c}{\bf{4th CSM shell}} \\
\cline{1-5}\\
$M_\mathrm{CSM,4}$ & $t_{i,4}$   &$\rho_\mathrm{in,4}$    &$\varepsilon_{\mathrm{fs},4}$  &$\varepsilon_{\mathrm{rs},4}$   \\
 ($M_\odot$)& (day) & ($10^{-14}$\,g\,cm$^{-3}$)\\
\cline{1-5}\\
$1.1_{-0.1}^{+0.1}$& $194_{-0.6}^{+0.4}$ &  $6.4_{-3}^{+4.8}$ & $0.041_{-0.012}^{+0.02}$ & $0.038_{-0.014}^{+0.02}$\\
\hline
\hline
\end{tabular}
\label{tab: Lum_fit}
\end{table}

\clearpage

\clearpage



\renewcommand{\tablename}{Supplementary Table} 
\renewcommand{\figurename}{Supplementary Figure} 
\setcounter{table}{0}
\setcounter{figure}{0}


\section{Supplementary Material}

\clearpage

\begin{table}
\center
\caption{Reference stars used for photometry calibration. The numbers in brackets are $1\sigma$ uncertainties.
\label{Table: referstar}}
\begin{tabular}{ccccccc}
\hline\hline
Star & $\alpha$(J2000)& $\delta$(J2000) & $B$ (mag) & $V$ (mag) & $R$ (mag) & $I$ (mag) \\
\hline
1 & $10^{\rm hr}18^{\rm m}44.11^{\rm s}$ & $46^{\circ}25'1.33''$ & 14.759(0.035) & 14.092(0.013) & 13.698(0.015) & 13.312(0.016)\\ 
2 & 10 18 44.99 & 46 32 11.98 & 19.257(0.037) & 17.748(0.015) & 16.896(0.016) & 15.755(0.016)\\ 
3 & 10 18 53.4 & 46 23 41.25 & 19.868(0.057) & 18.312(0.019) & 17.435(0.015) & 16.013(0.016)\\ 
4 & 10 18 55.41 & 46 22 48.51 & 16.085(0.039) & 15.053(0.014) & 14.461(0.016) & 13.917(0.016)\\ 
5 & 10 18 58.5 & 46 23 08.08 & 17.469(0.034) & 16.667(0.012) & 16.199(0.016) & 15.784(0.016)\\ 
6 & 10 19 0.11 & 46 27 20.11 & 19.177(0.035) & 18.214(0.013) & 17.659(0.015) & 17.164(0.016)\\ 
7 & 10 19 0.52 & 46 25 01.75 & 18.007(0.036) & 16.537(0.013) & 15.706(0.015) & 14.816(0.016)\\ 
8 & 10 19 13.21 & 46 30 59.5 & 17.761(0.034) & 16.23(0.016) & 15.366(0.017) & 14.197(0.017)\\ 
9 & 10 19 20.64 & 46 31 24.98 & 16.365(0.034) & 15.971(0.014) & 15.725(0.016) & 15.433(0.017)\\ 
10 & 10 19 26.18 & 46 24 04.76 & 19.73(0.036) & 18.177(0.014) & 17.302(0.016) & 16.048(0.017)\\ 
\hline
\end{tabular}
\end{table}

\clearpage

\begin{longtable}{lrrrrr}
\caption{$BVRI$ Photometry of SN~2017egm. The numbers in brackets are $1\sigma$ uncertainties in units of 0.001\,mag.\label{Table: phot}}\\
\hline\hline
\multicolumn{1}{c}{MJD} & \multicolumn{1}{c}{$B$ (mag)} & \multicolumn{1}{c}{$V$ (mag)} & \multicolumn{1}{c}{$R$ (mag)} & \multicolumn{1}{c}{$I$ (mag)} & \multicolumn{1}{c}{Telescope}  \\
\hline
\tiny
\endfirsthead
\caption{continued.}\\
\hline\hline
\multicolumn{1}{c}{MJD} & \multicolumn{1}{c}{$B$ (mag)} & \multicolumn{1}{c}{$V$ (mag)} & \multicolumn{1}{c}{$R$ (mag)} & \multicolumn{1}{c}{$I$ (mag)} & \multicolumn{1}{c}{Telescope}  \\
\hline
\endhead
\hline
\endfoot
57905.56 & 15.941(31) & 15.969(23) & 15.997(31) & 15.928(25) & TNT\\ 
57906.56 & -- & 15.773(20) & 15.830(20) & 15.922(24) & TNT\\ 
57908.53 & -- & -- & 15.748(12) & -- & TNT\\ 
57911.54 & 15.566(23) & 15.552(18) & 15.632(34) & 15.598(13) & TNT\\ 
57911.65 & 15.551(27) & 15.546(4) & -- & -- & 50BiN\\ 
57912.56 & 15.425(28) & 15.477(10) & 15.551(24) & 15.527(14) & TNT\\ 
57913.58 & 15.359(196) & 15.449(11) & -- & -- & 50BiN\\ 
57917.53 & 15.139(19) & 15.045(23) & 15.220(27) & 15.237(19) & TNT\\ 
57919.54 & 14.974(20) & 14.992(9) & 15.078(22) & 15.048(10) & TNT\\ 
57923.61 & 14.732(7) & 14.829(6) & -- & -- & 50BiN\\ 
57924.58 & 14.720(44) & 14.842(6) & -- & -- & 50BiN\\ 
57925.6 & 14.872(147) & 14.806(13) & -- & -- & 50BiN\\ 
57926.61 & 14.820(110) & 14.806(7) & -- & -- & 50BiN\\ 
57927.62 & 14.875(15) & 14.818(18) & -- & -- & 50BiN\\ 
57929.59 & 14.911(20) & 14.848(28) & -- & -- & 50BiN\\ 
57933.55 & -- & 14.931(46) & 14.849(26) & 14.965(44) & TNT\\ 
57933.59 & 14.922(36) & 14.933(20) & -- & -- & 50BiN\\ 
57940.59 & -- & 15.038(105) & -- & -- & 50BiN\\ 
57945.58 & 14.914(43) & 15.100(22) & -- & -- & 50BiN\\ 
57946.57 & 15.017(43) & -- & -- & -- & 50BiN\\ 
57947.58 & -- & -- & 14.945(17) & -- & 50BiN\\ 
58023.86 & 16.943(23) & 16.459(17) & 16.218(29) & 15.857(18) & TNT\\ 
58024.87 & 17.111(39) & 16.543(16) & 16.384(28) & 16.011(26) & TNT\\ 
58025.87 & 17.233(60) & 16.561(24) & 16.434(36) & 15.998(32) & TNT\\ 
58038.88 & 17.528(43) & 16.853(19) & 16.675(30) & 16.225(24) & TNT\\ 
58039.88 & 17.844(175) & 16.992(141) & 16.768(41) & 16.385(70) & TNT\\ 
58041.87 & 17.731(52) & 16.912(14) & 16.780(32) & 16.301(22) & TNT\\ 
58045.88 & 17.623(29) & 16.934(12) & 16.750(29) & 16.311(15) & TNT\\ 
58052.86 & 17.482(28) & 16.860(16) & 16.694(28) & 16.193(15) & TNT\\ 
58053.89 & 17.523(53) & 16.893(21) & 16.725(33) & 16.268(19) & TNT\\ 
58054.87 & 17.428(28) & 16.753(18) & 16.749(32) & 16.142(20) & TNT\\ 
58055.87 & 17.676(23) & 17.114(12) & 16.971(30) & 16.457(16) & TNT\\ 
58056.81 & 17.803(25) & 17.209(22) & 17.098(30) & 16.512(19) & TNT\\ 
58059.87 & 18.212(51) & 17.637(30) & 17.452(42) & 16.926(26) & TNT\\ 
58060.86 & 18.140(67) & 17.824(48) & 17.622(48) & 17.068(30) & TNT\\ 
58062.9 & -- & 17.951(85) & 17.968(85) & 17.128(62) & TNT\\ 
58064.02 & 18.468(173) & 18.265(63) & 18.024(108) & 17.399(54) & AZT\\ 
58064.9 & 18.903(90) & 18.257(46) & 18.142(61) & 17.533(55) & TNT\\ 
58065.01 & 18.576(143) & 18.260(41) & 18.532(118) & 17.551(74) & AZT\\ 
58065.99 & 19.096(165) & 18.387(21) & 18.434(64) & 17.658(33) & AZT\\ 
58067.89 & 19.319(74) & 18.665(56) & 18.457(46) & 17.701(32) & TNT\\ 
58068.04 & 19.791(219) & 18.672(25) & 18.489(73) & 17.774(59) & AZT\\ 
58069.03 & 19.736(86) & 18.777(59) & 18.538(78) & 17.794(35) & AZT\\ 
58070.84 & 19.504(71) & 19.021(50) & 18.641(51) & 17.948(42) & TNT\\ 
58071.84 & 19.653(89) & 18.987(51) & 18.627(59) & 17.979(53) & TNT\\ 
58073.82 & 19.767(105) & 19.146(67) & 18.945(75) & 18.122(58) & TNT\\ 
58078.89 & -- & 19.307(90) & 18.846(78) & 18.187(62) & TNT\\ 
58085.82 & -- & 19.773(99) & 18.878(75) & 18.139(53) & TNT\\ 
58086.76 & -- & 19.209(75) & 18.760(86) & 18.187(61) & TNT\\ 
58087.8 & 19.983(108) & 19.452(70) & 18.842(65) & 18.007(38) & TNT\\ 
58088.77 & -- & -- & 18.600(113) & 17.913(76) & TNT\\ 
58089.79 & -- & -- & 18.503(103) & 17.937(74) & TNT\\ 
58096.91 & 19.818(162) & 18.796(50) & 18.462(58) & 17.691(29) & TNT\\ 
58098.74 & 19.865(118) & 18.854(46) & 18.469(53) & 17.690(30) & TNT\\ 
58099.81 & 19.854(76) & 19.049(38) & 18.633(57) & 17.759(26) & TNT\\ 
58102.79 & 19.636(142) & 19.141(65) & 18.844(54) & 17.976(45) & TNT\\ 
58105.83 & 19.970(98) & -- & 18.948(182) & -- & TNT\\ 
58106.95 & 20.123(236) & 19.277(48) & 18.966(70) & 18.208(56) & AZT\\ 
58127.9 & 19.749(187) & 19.174(106) & 18.594(111) & 18.016(40) & AZT\\ 
58132.99 & 19.904(88) & 19.074(59) & 18.756(76) & 17.981(52) & AZT\\ 
58134.75 & 19.722(94) & 19.224(78) & 18.637(52) & 17.827(39) & TNT\\ 
58135.88 & 19.880(325) & 19.039(89) & 18.849(86) & 17.971(43) & AZT\\ 
58136.71 & 19.730(67) & 19.086(40) & 18.658(52) & 17.879(38) & TNT\\ 
58138.95 & 19.717(81) & 18.931(39) & 18.704(86) & 17.845(67) & AZT\\ 
58139.92 & 19.756(158) & 18.860(48) & 18.664(76) & 17.819(31) & AZT\\ 
58140.93 & 19.915(173) & 18.853(48) & 18.742(76) & 17.815(23) & AZT\\ 
58143.85 & 19.680(111) & 18.899(46) & 18.698(75) & 17.849(35) & AZT\\ 
58148.78 & -- & -- & 18.233(83) & 17.766(61) & TNT\\ 
58150.8 & -- & -- & -- & 17.868(64) & TNT\\ 
58152.72 & 19.470(126) & 18.751(67) & 18.598(54) & 17.772(33) & TNT\\ 
58152.92 & 19.628(92) & 18.828(39) & 18.663(69) & 17.897(59) & AZT\\ 
58155.01 & 20.176(191) & 18.796(46) & 18.731(76) & 17.817(21) & AZT\\ 
58156.99 & 20.013(117) & 18.932(31) & 18.806(74) & 17.968(28) & AZT\\ 
58162.65 & 19.557(59) & 19.186(51) & 19.016(59) & 17.988(40) & TNT\\ 
58167.9 & 20.428(127) & 19.688(83) & 19.655(96) & 18.653(73) & AZT\\ 
58170.9 & 20.681(221) & 19.941(54) & 19.785(107) & 18.870(72) & AZT\\ 
58171.88 & 20.696(182) & 20.028(61) & 20.014(101) & 19.046(44) & AZT\\ 
58193.9 & -- & -- & 20.657(34) & -- & AZT\\ 
58197.69 & -- & -- & 20.697(224) & 19.632(123) & AZT\\ 
58211.81 & -- & -- & 21.130(190) & 19.717(78) & AZT\\ 
58212.76 & -- & -- & -- & 19.547(62) & AZT\\ 
58234.69 & -- & -- & -- & 19.961(99) & AZT\\ 
58235.7 & -- & -- & 21.271(196) & 19.832(85) & AZT\\ 
58241.73 & -- & -- & -- & 20.109(141) & AZT\\ 
58242.79 & -- & -- & 21.802(246) & 19.991(85) & AZT\\ 
58245.68 & -- & -- & -- & 19.949(85) & AZT\\ 
58264.74 & -- & -- & -- & 20.216(144) & AZT\\ 
58269.71 & -- & -- & -- & 19.986(98) & AZT\\ 
58270.67 & -- & -- & -- & 20.268(136) & AZT\\ 
58275.72 & -- & -- & 21.623(231) & 20.170(81) & AZT\\ 

\end{longtable}

\begin{table}
\center
\caption{{\it Swift} Photometry of SN~2017egm. The numbers in brackets are $1\sigma$ uncertainties in units of 0.001\,mag.}
\begin{tabular}{ccccccc}
\hline\hline
 MJD & $UVW2$ (mag) & $UVM2$ (mag) & $UVW1$ (mag) & $U$ (mag) &  $B$ (mag) & $V$ (mag) \\
\hline
57906.3 & 14.239(46) & 13.972(46) & 14.133(46) & 14.461(56) & 15.875(72) & 15.891(104)\\ 
57906.4 & 14.172(47) & 13.965(49) & 14.040(46) & -- & 15.803(77) & 15.877(119)\\ 
57908.2 & 14.299(56) & 14.092(58) & 14.144(54) & 14.399(50) & 15.629(72) & 15.862(156)\\ 
57909.1 & 14.185(50) & 13.917(50) & 14.048(49) & 14.421(50) & 15.624(72) & 15.764(132)\\ 
57912.1 & 14.088(60) & -- & -- & 14.158(50) & -- & --\\ 
57912.4 & 14.059(59) & -- & -- & 14.140(50) & -- & --\\ 
57912.7 & -- & -- & -- & 14.067(48) & -- & --\\ 
57912.8 & 14.034(59) & -- & -- & -- & -- & --\\ 
57914.5 & 13.943(52) & 13.636(51) & 13.707(50) & 13.912(50) & 15.258(68) & 15.407(99)\\ 
57916.0 & 13.924(49) & 13.650(49) & 13.664(48) & 13.850(48) & 15.191(63) & 15.161(74)\\ 
57919.2 & 13.865(49) & 13.549(48) & 13.485(46) & 13.626(46) & 14.962(58) & 15.090(73)\\ 
57920.3 & 13.734(47) & 13.420(46) & 13.368(45) & 13.562(46) & 14.969(61) & 14.945(70)\\ 
57922.2 & 13.731(47) & 13.363(46) & 13.303(45) & 13.449(45) & 14.790(55) & 14.834(68)\\ 
57924.2 & 13.842(48) & 13.475(47) & 13.342(45) & 13.416(44) & 14.750(54) & 14.828(67)\\ 
57925.7 & 13.946(52) & -- & -- & -- & -- & --\\ 
57926.7 & 14.016(51) & 13.606(49) & 13.434(46) & 13.448(45) & 14.804(56) & 14.786(68)\\ 
57928.7 & 14.198(53) & 13.771(50) & 13.588(47) & 13.548(46) & 14.825(56) & 14.845(68)\\ 
57932.6 & 14.538(59) & 14.147(56) & 13.853(50) & 13.623(46) & 14.921(57) & 14.941(70)\\ 
57934.7 & -- & -- & 13.981(51) & -- & -- & --\\ 
57934.8 & 14.704(63) & 14.320(60) & -- & 13.695(47) & 14.972(59) & 14.892(68)\\ 
57937.4 & 14.813(68) & 14.504(61) & 14.123(54) & 13.780(48) & 14.995(59) & 14.941(72)\\ 
57938.1 & 14.890(71) & 14.581(62) & 14.188(56) & 13.816(48) & 15.071(64) & 14.869(71)\\ 
58026.3 & >18.471 & >18.532 & 18.061(283) & 16.898(143) & -- & --\\ 
58031.0 & >18.399 & >18.185 & 17.962(313) & 17.103(172) & -- & --\\ 
58031.4 & >18.568 & >18.626 & 18.017(276) & 17.011(149) & 17.532(143) & 16.676(141)\\ 
58036.3 & >18.647 & >18.537 & 18.361(350) & 17.271(161) & -- & --\\ 
58041.4 & >18.644 & >18.522 & >18.493 & 17.403(178) & -- & --\\ 
58046.3 & >18.51 & >18.377 & >18.357 & 17.347(183) & -- & --\\ 
58051.4 & >18.651 & >18.572 & >18.536 & 17.279(160) & -- & --\\ 
58056.0 & >18.581 & >18.426 & 18.304(354) & 17.578(210) & -- & --\\ 
58058.9 & >18.535 & >18.6 & >18.422 & 17.926(288) & 17.817(180) & 17.156(204)\\ 
58083.4 & >18.68 & >18.588 & >18.55 & >18.544 & -- & --\\ 
58096.3 & -- & -- & -- & >18.619 & -- & >18.261\\ 
58130.5 & -- & -- & -- & -- & -- & >18.303\\ 
58295.3 & >18.656 & -- & -- & -- & -- & --\\ 
\hline
\end{tabular}
\label{Table: phot_swift}
\end{table}

\begin{table*}
\center
\caption{Journal of Spectroscopic Observations of SN~2017egm.}
\begin{tabular}{cccccccc}
\hline
\hline
UT & MJD & Phase$^{a}$ & Instrument & Exp. time (s) & Range (\AA)  \\
\hline
2017-05-26 &  57899.5 & $-$26.4 & XLT/BFOSC &2700  & 3721--8781  \\
2017-05-30 &  57903.6 & $-$22.5 & XLT/BFOSC  &3000  & 3721--8794  \\
2017-05-31 &  57904.5 & $-$21.5 & XLT/BFOSC  &3000  & 3724--8793  \\
2017-06-05 &  57909.6 & $-$16.6 & LJT/YFOSC  &1600  & 3501--9170  \\
2017-06-07 &  57911.6 & $-$14.7 & XLT/BFOSC &3000  & 3711--8780  \\
2017-06-17 &  57921.1 & $-$5.4 &  P200/DBSP  &  600   &  3800--10,000 \\
2017-06-17 &  57921.5 & $-$5 &   XLT/BFOSC&  2700 &  3715--8660  \\
2017-06-22 &  57926.2 & $-0.5$ &  P200/DBSP     &   600   & 3800--10,000\\
2017-09-27 & 58023.5 & 93.9     &   Lick-3m/Kast  &  2130  &   3630--10,680 \\
2017-10-19  &  58045.5 & 115.3    &    Lick-3m/Kast  &  1230  &   3622--10,670 \\
2017-10-19 &  58045.9 & 115.6 &   XLT/BFOSC & 3000 &3730--8822 \\
2017-10-25  &  58051.5 &  121.1    & Lick-3m/Kast  & 2130   &   3620--10,680\\
2017-10-26 &  58052.4 & 122   &  P200/DBSP   &    1200    &3500--9000   \\
2017-10-30  &  58056.5 &  126    & Lick-3m/Kast  & 3030   &   3622--10,712\\
2017-11-15 &  58072.6 & 141.5 & Keck-I/LRIS   &   900   &  3072--10,293   \\
2017-11-16 &  58073.6 & 142.5 & Keck-I/LRIS   &   1165   &  3132--10,238   \\
2017-11-21  &  58078.5 &  147.4    & Lick-3m/Kast  &  2010  &   3632--10,710\\
2017-12-12 &  58099.5  &  167.5    &Lick-3m/Kast  &  3630   &  3630--10,680 \\
2017-12-18&   58105.5  &  173.5    & Lick-3m/Kast  &  3630  &   3632--10,716\\
2018-01-08 &  58126 & 193.4 &    Keck-I/MOSFIRE&   1074     & 9751--11,240\\
2018-01-08 &  58126 & 193.4 &    Keck-I/MOSFIRE&    954   & 11,530--13,518\\
2018-01-13 &  58131.5 & 198.7    & Lick-3m/Kast  &  3630  &   3630--10,680\\
2018-01-13 &  58131.5 & 198.7 & Keck-I/LRIS   &   1760   &  3066--10,264   \\
2018-05-10 &  58248.4 & 312.1 & Keck-I/LRIS   &   1757   &  3149--10,237 \\
2018-07-13 &  58312.3 & 374.1 & Keck-I/LRIS   &   2550   &  3099--10,287 \\
\hline
host galaxy\\
\hline
2018-01-13 &  58131.5 & 198.7 & Keck-I/LRIS   &   1760   &  3066--10,264   \\\hline
\end{tabular}
\\$^{a}${Rest-frame days with respect to the epoch of maximum $V$ brightness.}
\label{Table: spec_Journal}
\end{table*}
\clearpage

\begin{table}
\center
\caption{Free Parameters and priors of the MCSIRD Model. Here $t_\mathrm{exp}$ is the explosion date from the maximum light; $x_0$ is the dimensionless transitional radius between inner and outer part of ejecta; $E_{\mathrm{SN}}$ and $M_{\mathrm{ej}}$ are the kinetic energy and mass of the SN ejecta; $M_\mathrm{Ni}$ is the newly-synthesised $^{56}$Ni mass; $t_{i,k}$ is the initial time of $k$-th interaction; $M_\mathrm{CSM,k}$ is the mass of $k$-th CSM shell, $\rho_\mathrm{in,k}$ is the CSM density at the inner radius ($R_\mathrm{in,k}$), and $\varepsilon_{\mathrm{fs},k}$ and $\varepsilon_{\mathrm{rs},k}$ are the efficiencies of converting forward and reverse shock energy to radiation. Here the subscript $k$ ($=1,2,3,4$) denotes the sequential number of CSM shell.}
\begin{tabular}{cccccccc}
\hline\hline
Parameter &  Unit & Prior   & Min   &Max \\\hline
$t_\mathrm{exp}$ &day &  Flat  & -100   &-32 \\
$x_0$ &-- &  Flat &   $0.1$  &   $1$  \\
$E_{\mathrm{SN},1}$&$10^{51}$ erg&  Log-Flat &   $0.01$  &   $100$  \\
$M_{\mathrm{ej},1}$&$M_\odot$ & Log-Flat & 0.01    &100 \\
$M_\mathrm{Ni}$  &$M_\odot$ & Log-Flat & 0  &10  \\
$R_\mathrm{in,1}$& $10^{15}$ cm & Log-Flat & 0.0001    &1000 \\
$t_{i,2}$& day & Flat & 0 & 123\\
$t_{i,3}$& day & Flat & 123& 167 \\
$t_{i,4}$& day & Flat & 167& 219\\
$M_\mathrm{CSM,k}$ &$M_\odot$ & Log-Flat & 0.01    &100 \\
$\rho_\mathrm{in,k}$ & $10^{-14}$ g\,cm$^{-3}$ & Log-Flat & 0.0001    &10000 \\
$\varepsilon_{\mathrm{fs},k}$&-- & Log-Flat & 0.01  &1\\
$\varepsilon_{\mathrm{rs},k}$&-- & Log-Flat & 0.01  &1\\
\hline
\hline
\end{tabular}
\label{tab: priors_fit}
\end{table}

\begin{figure}
\center
\includegraphics[angle=0,width=1\textwidth]{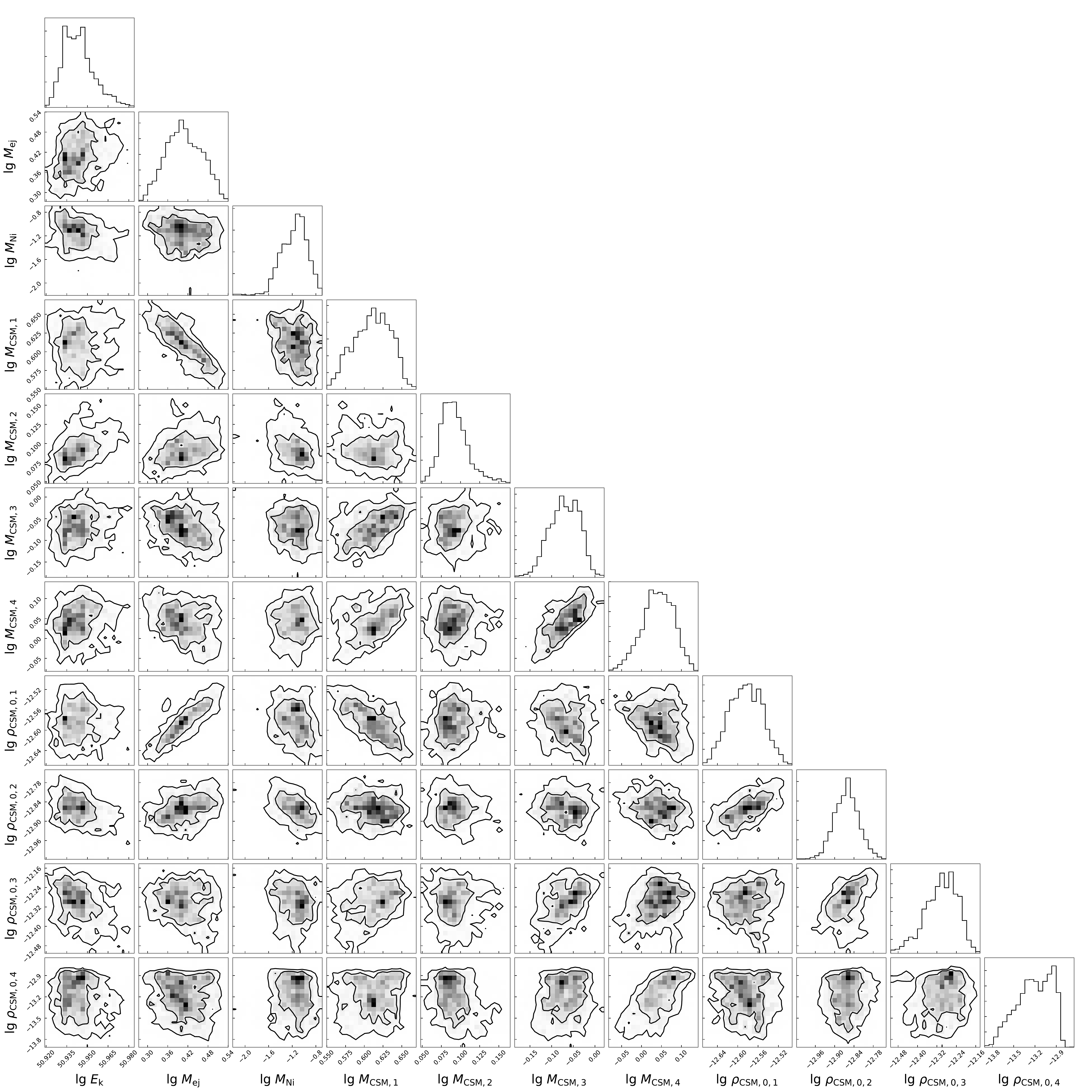}
\caption{Posteriors for some key parameters inferred for SN~2017egm using the MCSIRD model. The mean and $1\sigma$ uncertainties of fitting parameters are presented in Extended Data Table~\ref{tab: Lum_fit}.}
\label{fig: mcmc}
\end{figure}
\clearpage



\end{document}